\documentclass[
reprint,aps,
amsmath,amssymb,
prx
]{revtex4-2}

\usepackage{times}

\usepackage{mathtools}
\usepackage{xcolor}
\usepackage{graphicx}
\usepackage{bm}
\usepackage{float}

\mathchardef\mhyphen="2D

\begin{document}

\title{Emergent predictability in microbial ecosystems}

\author{Jacob Moran}
\affiliation{Department of Physics, Washington University in St Louis, St Louis, MO, USA}
\affiliation{Department of Biophysics, University of Michigan, Ann Arbor, MI, USA}
\author{Lucas C. Graham}
\affiliation{Department of Physics, Washington University in St Louis, St Louis, MO, USA}
\author{Mikhail Tikhonov}
\email{tikhonov@wustl.edu}
\affiliation{Department of Physics, Washington University in St Louis, St Louis, MO, USA}

\date{\today}


\begin{abstract}
Microbial ecosystems exhibit a surprising amount of functionally relevant diversity at all levels of taxonomic resolution, presenting a significant challenge for most modeling frameworks. A long-standing hope of theoretical ecology is that some patterns might persist despite community complexity -- or perhaps even emerge because of it. A deeper understanding of such ``emergent simplicity'' could enable new approaches for predicting the behaviors of the complex ecosystems in nature. However, the concept remains partly intuitive with no consistent definition, and most empirical examples described so far afford limited predictive power. Here, we propose an information-theoretic framework for defining and quantifying emergent simplicity in empirical data based on the ability of coarsened descriptions to predict community-level functional properties. Applying this framework to two published datasets, we demonstrate that all five properties measured across both experiments exhibit robust evidence of what we define as ``emergent predictability'': surprisingly, as community richness increases, simple compositional descriptions become more predictive. We show that standard theoretical models of high-diversity ecosystems fail to recapitulate this behavior. This is in contrast to simple self-averaging, which is well-understood and generic across models. We propose that, counterintuitively, emergent predictability arises when physiological or environmental feedbacks \textit{oppose} statistical self-averaging along some axes of community variation. As a result, these axes of variation become increasingly predictive of community function at high richness. We demonstrate this mechanism in a minimal model, and argue that explaining and leveraging emergent predictability will require integrating large-$N$ theoretical models with a minimal notion of physiology, which the dominant modeling frameworks currently omit.
\end{abstract}

\maketitle
Microbial communities perform key functions in a wide range of contexts, including human health, agriculture, and the global climate. Predicting community function from its composition is a key goal of microbial ecology. Recent advances in high-throughput sequencing allowed compositional information to be obtained in high resolution, and revealed that natural communities consistently exhibit high richness (number of coexisting taxa), a surprising diversity at all levels of resolution, high functional redundancy, and a lack of reproducibility in even similar conditions. These characteristics pose significant challenges for making quantitative predictions, which are necessary for applications in industry and medicine.

A long-standing hope of theoretical ecology has been that, despite this complexity, some patterns might nevertheless be predictable~\cite{louca_high_2016,ward2017annual,vila2023metabolic}. The search for emergent regularities in complex ecosystems is as old as ecology itself. For example, classical work uncovered reproducible trophic structures across taxonomically divergent communities~\cite{heatwole1972trophic} or a consistent ratio of predator species to prey species~\cite{cohen1977ratio,briand1984community}. More recently, moving beyond the patterns persisting despite high ecosystem diversity, it was proposed that some regularities could perhaps arise \emph{because} of it~\cite{goldford_emergent_2018,fant_eco-evolutionary_2021,arya2023sparsity}, coining the term ``emergent simplicity''~\cite{goldford_emergent_2018,estrela_functional_2022,mcgrady1997biodiversity,naeem1997biodiversity,carrara2015experimental}. This notion is at the foundation of the interface between statistical physics and the so-called large-$N$ ecology, explaining species abundance distributions and other patterns as some parameter (e.g., the number of species) becomes large~\cite{advani_statistical_2018,marsland_minimal_2020,cui2021diverse,louca_decoupling_2016,louca_high_2016,fant_eco-evolutionary_2021,o2020beyond,odwyer_backbones_2015}.

The idea of emergent simplicity is very appealing, suggesting that functional predictions could perhaps be made directly in the complex natural context~\cite{marsland_minimal_2020,bergelson_functional_2021}. However, so far this hope remained out of reach. In the absence of a consistent definition, it is unclear precisely what forms of  ``emergent simplicity'' should be seen as surprising or useful. Many reported instances fall under the umbrella of functional convergence, focusing on reproducibility rather than prediction~\cite{heatwole1972trophic,cohen1977ratio,briand1984community,datta2016microbial,goldford_emergent_2018,estrela_functional_2022,pollak_public_2021,pontrelli2022metabolic,mcgrady1997biodiversity,naeem1997biodiversity,carrara2015experimental}. Such reproducibility can be difficult to distinguish from simple self-averaging or the central limit theorem~\cite{marsland2020minimal,ho2024challenges} and offers limited predictive power. As an illustration, consider the statement: ``if the soil ecosystem is diverse enough, the rate of process $P$ will be $\approx\!\!X$ per gram of biomass''. This blanket expectation would likely be somewhat na\"ive, as  empirical observations of reproducibility are often approximate or limited to simplified conditions~\cite{silverstein2023metabolic}. But even if it were true, it would leave key questions unanswered, such as how the process rate would be affected by a perturbation, or what intervention would help increase it. As a result, observing emergent regularities of this kind has generally not led to improved functional predictions.

Understanding and harnessing emergent simplicity could transform our ability to predict and control complex natural communities. However, to realize this potential we need to make this concept well-defined, measurable, and focused on $\textit{prediction}$ rather than reproducibility. Here, we introduce a theoretical framework to explicitly quantify the predictive power of coarsened compositional descriptions in ecological data. This allows us to convert the aspiration of emergent simplicity into a testable quantitative hypothesis. Applying our framework to two published datasets on lab-assembled ecosystems, we show that, remarkably, for all 5 properties measured across both experiments, the predictive power of coarse descriptions indeed improves with community richness. We further show that standard theoretical models of high-diversity ecosystems fail to recapitulate this behavior. This is in contrast to simple self-averaging, which is well-understood and generic across models. We propose that, counterintuitively, emergent predictability arises when physiological or environmental feedbacks \textit{oppose} statistical self-averaging along some axes of community variation. As a result, these axes of variation can become increasingly predictive of community function at high richness. We demonstrate this mechanism in a minimal model, and argue that explaining and leveraging emergent predictability will require integrating large-$N$ theoretical models with a minimal notion of physiology, which the dominant modeling frameworks currently omit.

\section{The theoretical framework}
\subsection{A model experimental setup}\label{sec:exp_setup}
\begin{figure*}[]
\centering
 \includegraphics[width=\linewidth]{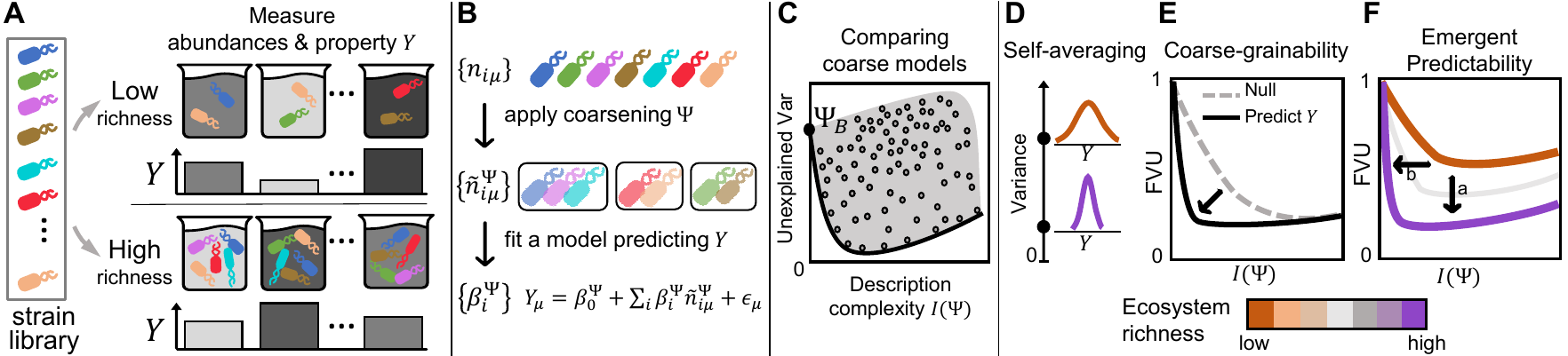}
\caption{
\textbf{A quantitative framework for interrogating emergent simplicity in data.} \textbf{A:} A general experimental protocol with community richness (number of strains) as a control parameter. A fixed library of strains is subsampled to assemble communities of low richness (top) or high richness (bottom). Community composition (strain abundances, $n_i$) and some property of interest $Y$ are measured in each community. \textbf{B:} Our approach will be to train models predicting $Y$ based on  combined abundances of groups of taxa. Each choice of a  \textit{coarsening} $\Psi$ defines a regression model (coefficients $\{\beta_i^\Psi\}$), whose performance we will evaluate. \textbf{C:} For each coarsening $\Psi$, we can evaluate the \textit{description complexity} $I(\Psi)$ (the extent of coarsening; see text) and the out-of-sample prediction error, quantified as the variance unexplained by the corresponding regression model. The cartoon illustrates the expected general shape of the scatter plot in these axes, evaluated for all possible coarsenings $\Psi$ (points). Our focus is on the shape of the Pareto front (thick line), characterizing the best predictive power achievable by models of a given complexity.
The behavior of the Pareto fronts defines three distinct (but not mutually exclusive) forms of ``emergent simplicity'': \textbf{D:} self-averaging, when higher community richness leads to a tighter distribution of property $Y$; \textbf{E:} coarse-grainability, when coarsened descriptions are surprisingly predictive (compared to a null model), at fixed richness; or \textbf{F:} emergent predictability, when the predictive power of coarsened descriptions improves with ecosystem richness.
\label{fig:1}}
\end{figure*}

For concreteness, consider the experimental setup depicted in Fig.~\ref{fig:1}A. Let $\mathcal L$ be a fixed library of $S$ strains, indexed by $i$. Subsets of these strains are sampled from this library to inoculate $N$ communities, indexed by $\mu$ and defined by a binary matrix $b_{i\mu}$ identifying whether strain $i$ is included in community $\mu$. We will assume that community richness $R_\mu\equiv \sum_i b_{i\mu}$ is kept consistent across the dataset; this richness will become our key control parameter for comparisons (a low-richness dataset \textit{vs.}\ a high-richness dataset). Once the communities are assembled, we measure their microscopic composition (abundance of each strain $n_{i\mu}$) as well as some community-level property of interest $Y_\mu$ that we seek to predict (e.g., production of a metabolite). We will not assume that the communities reach any notion of a steady state, merely that some assembly protocol is followed consistently in each trial. Our goal is to quantify the \emph{predictive power of simple (coarsened) descriptions} in ecological datasets of this kind, and then use empirical data to ask how this predictive power changes with community richness.

\subsection{The candidate `simple models'}\label{candidate_models}
Our approach is motivated by an empirical example, namely wastewater treatment bioreactors that use microbes to digest organic waste into methane. These complex communities assemble semi-naturally and can harbor over 1000 taxa~\cite{jeglot_microbiome_2021}. The rate of digestion (the amount of methane produced) can vary widely~\cite{korres2013variation,naik2014factors}, but, remarkably, can be largely explained by simple models coarse-graining this vast diversity into just 4 functional classes (acidogens, propionate-degrading acetogens, butyrate-degrading acetogens, and methanogens~\cite{bertacchi_mathematical_2021}).

Motivated by this example, throughout this work, our candidate simple models will be indexed by the choice of a \emph{coarsening} $\Psi$, which maps individual strains into a smaller number of classes: $[1\dots S] \xrightarrow{\Psi} [1\dots K^\Psi]$ with $K^\Psi\leq S$ (Fig.~\ref{fig:1}B). Some natural choices of coarsening $\Psi$ might  be to group strains based on their taxonomy or key traits, but we will allow any choice of $\Psi$ as valid, and define \emph{description complexity} $I(\Psi)$ as the mutual information between the label of the coarse-grained class and the identity of the microscopic strain:
\[I(\Psi)=-\sum_{j=1}^{K^\Psi} \frac{S^\Psi_j}{S} \log \frac{S^\Psi_j}{S}.\]
Here $S_j^\Psi$ is the number of strains that $\Psi$ assigns to class $j$; $K^\Psi$ is the total number of classes; and $S$ is the total number of strains in the library. Our definition of $I(\Psi)$ implements the intuition that description complexity should grow with the number of classes it distinguishes, but generalizes it to groups of unequal size (for additional discussion, see Supplementary Text, section~\ref{si:descr_compl}). This metric $I(\Psi)$ reaches its maximum for the fully microscopic description, when each strain is its own class: $K^\Psi=S$ and $I(\Psi)=\log S$.

For any coarsening $\Psi$, we will assess the prediction error of the best model from a specified \emph{model class} $\mathcal M$, taking as input the combined abundances $\tilde n^\Psi_{j\mu}=\sum_{\Psi(i)=j}n_{i\mu}$ and predicting $Y$. This approach echoes other recent attempts to relate community-level properties to coarse-grained compositional features~\cite{shan2023annotation,moran_defining_2022}. The model class $\mathcal M$ could be mechanistically motivated (for instance, the class of Generalized Lotka-Volterra models predicting a dynamical property like future abundance of a focal species), or some general-purpose ansatz (e.g.,\ a neural network). In this work, we will pick a particularly simple class of models, namely linear regressions $Y_\mu=\beta^\Psi_0+\sum_i \beta^\Psi_i \tilde n^{\Psi}_{i\mu}+\epsilon_\mu$ (where $\epsilon_\mu$ is the residual noise). This choice is motivated by two considerations. First, we will want to apply our framework to data, and as we will see, the limitations of existing datasets make it preferable to pick a model class trainable on fewest data points. Second, it was recently shown that simple regressions can in fact be surprisingly predictive of ecosystem function~\cite{skwara2023statistically,arya2023sparsity}, motivating us to proceed.

In summary, the candidate ``simple models'' we will consider are indexed by the choice of a coarsening $\Psi$. Each $\Psi$ defines a set of coarsened variables ($\{\tilde n^{\Psi}_{i\mu}\}$), which allow us to train an instance of a linear regression model (coefficients $\beta^{\Psi}_i$), whose prediction error we will evaluate. It is worth noting that a common approach to vary modeling complexity is to keep the variables fixed (e.g. ``foxes'', ``rabbits''), but change the model class (e.g, a non-interacting model; adding quadratic Lotka-Volterra interactions; then higher-order terms; etc.). In contrast, the defining feature of our approach is to fix the model class, and change the variables (vary the extent of taxon grouping into putative functional classes).

\subsection{Quantifying the performance of simple models}
For each candidate ``simple model'' (coefficients $\beta^{\Psi}_i$) indexed by the coarsening $\Psi$, we can quantify its  prediction error, computed on held-out data. The well-established bias-variance tradeoff from statistical learning theory indicates that if we could evaluate and plot the prediction error (the variance left unexplained)
against the description complexity $I(\Psi)$ for \emph{all} possible coarsenings $\Psi$, the result would be expected to look like Fig.~\ref{fig:1}C. Each point in the point cloud corresponds to one choice of $\Psi$, and describes the predictive power of one candidate simple model. The left-most point $\Psi_B$ is the least predictive, coarsest description that groups all strains together ($I(\Psi_B)=0$). If no compositional information is available---only the total biomass---the unexplained variance of $Y$ is close to its total variance. (For properties that scale with biomass, it might be more meaningful to seek to predict their  value per unit biomass; see Supplementary Text, section~\ref{si:normalization}.)  As the information content of a description increases, better predictions become possible. Models that provide the lowest prediction error for a given complexity constitute the Pareto front (thick line). The Pareto front continues to trend downwards until descriptions become too complex to be fit on limited data (overfitting), and out-of-sample prediction error starts increasing (Fig.~\ref{fig:1}C). From now on, our focus will be on the behavior of this Pareto front.

\subsection{Distinguishing self-averaging, coarse-grainability and emergent predictability}
The intuitive notion of ``emergent simplicity'' can be codified in three distinct ways. These are important to distinguish, since as we will see, they have very different properties.

The simplest, and most familiar, version of emergent simplicity is self-averaging. As community richness increases, one might observe the distribution of $Y_\mu$ become increasingly tight. Much previous work indicates that properties of high-richness communities may indeed exhibit such ``functional convergence.''~\cite{heatwole1972trophic,cohen1977ratio,briand1984community,datta2016microbial,goldford_emergent_2018,estrela_functional_2022,pollak_public_2021,pontrelli2022metabolic,mcgrady1997biodiversity,naeem1997biodiversity,carrara2015experimental,fant_eco-evolutionary_2021}
We can think of this width as the prediction error of a particularly trivial model which predicts a constant value, such that the unexplained variance is synonymous with the total variance.

In practice, the predictions that interest us are rarely of this ``zero-information'' kind. Instead, we usually seek to predict properties of a community based on its composition. To define notions of simplicity that go beyond simple self-averaging, it is convenient to normalize away the overall width of the distribution of $Y$, and think of the \textit{fraction} of unexplained variance (FVU), defined as the mean square error of prediction divided by the variance of $Y$.

In these normalized units, the Pareto front descends from approximately 1 (no compositional information; worst prediction) to the best prediction attainable with this model class (in the limit of infinite data---neglecting overfitting---this would correspond to the full microscopic description). How quickly it descends encodes the notion of ``coarse-grainability''. For some properties, we may find that a surprisingly coarse description already enables good predictions; this form of simplicity has also been widely reported~\cite{Geisert2022MiceGut,Lee2025TheBeast,Rao2021MultiKingdomGLV,Shahin2023EMBED}. To rigorously define this concept, it is essential to quantify the word ``surprisingly,'' defining a null expectation to which the shape of this curve should be compared. Our framework will enable us to do so, allowing us to examine ``coarse-grainability'' of ecological properties in empirical data.

Finally, we can ask what happens to the Pareto front when we change ecosystem richness, and compare the fronts constructed for a dataset restricted to only high-richness examples \textit{vs.}\ only low-richness examples, for the same property $Y$ (Fig.~\ref{fig:1}A). A consistent shift of the Pareto front would indicate that, as richness is increased, models of comparable complexity become increasingly predictive (Fig.~\ref{fig:1}F, arrow ``a''), and/or that same prediction accuracy can be achieved by increasingly coarse descriptions (Fig.~\ref{fig:1}F, arrow ``b''). While the two directions of shift have different verbal interpretations, in general they cannot be formally distinguished, and we will instead quantify ``emergent predictability'' by measuring the area contained between the two Pareto fronts (see Methods).

The rest of the paper can be summarized in two claims: (1) Self-averaging, coarse-grainability and emergent predictability are all observed in empirical data. However, (2) the three phenomena differ dramatically in their origins, properties and the attention they deserve. Today, the broad usage of the phrase ``emergent simplicity'' came to encompass any observation of something simple in a complex community context. In this sense, all three phenomena just defined could be collated under this umbrella term. However, as we will show, self-averaging is generic and ubiquitous; on its own, it is neither surprising nor special. Coarse-grainability does not come ``for free'' (from diversity alone), but can be a straightforward consequence of the fact that some species are more similar than others. While missing from some widely used models that omit such structure, its empirical observation should also not be seen as surprising, unless the observation comes with a quantitative argument that the observed coarse-grainability is stronger than expected. In contrast, emergent predictability is missing from both random and structured theoretical models -- and yet, as we will see, it is strongly supported by data. This is the phenomenon that we propose should be seen as ``emergent simplicity'' \textit{sensu stricto}. Its empirical observation is surprising and demands a theoretical explanation, and we will present evidence supporting a candidate mechanism for its emergence.

\section{Results}
\subsection{Data selection}
To apply our framework, we need a dataset satisfying several conditions. First, it must follow the combinatorial assembly protocol of Fig.~\ref{fig:1}A, with communities assembled from a fixed pool of strains. Surveying the literature, we identified 9 candidate datasets (across both microbial and non-microbial contexts; see Supp. Table~\ref{tab:table1}). However, for our purposes it is essential that the assayed communities span a range of richness, so that at least two tiers of community diversity can be defined and compared. Additionally, if $N$ is the number of samples and $S$ the number of strains in the pool, $N/S$ should be sufficiently large (for each richness tier) to allow training the model from our chosen model class. Our choice of linear regressions as model class makes this last condition the most permissive and identifies two datasets whose structure allows our question to be asked. These are the data of Clark \textit{et~al.}~\cite{clark_design_2021}, who assayed communities of 1-25 gut-derived taxa and measured the production of butyrate, acetate, lactate and succinate; and the work of Kehe \textit{et~al.}~\cite{kehe_massively_2019}, who used a droplet-based platform to screen communities of 1-14 isolates, measuring the abundance of a model plant symbiont \textit{H.~frisingense} as the property of interest. For each dataset, we resolve three tiers of community richness: low richness, $1-5$ strains; middle richness, $10-15$ strains; and high richness, $21-25$ strains. Together, these two datasets provide measurements of five properties we can examine for evidence of emergent simplicity. To construct the information-prediction Pareto fronts defined above, we use a Metropolis-Hastings algorithm to search the space of possible coarsenings (for details, see Methods).

\subsection{Self-averaging is generic and ubiquitous}
\begin{figure}[t!]
\centering
\includegraphics[width=0.8\linewidth]{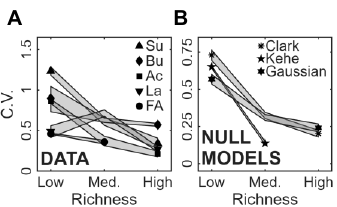}
\caption{
\textbf{Self-averaging is generic.}
\textbf{A:} Coefficient of variation (CV) for each of the properties measured in Refs.~\cite{clark_design_2021} and~\cite{kehe_massively_2019}, computed across communities of a given richness tier. The decreasing trend indicates self-averaging.  Legend labels: Bu~$=$~Butyrate, Ac~$=$~Acetate, La~$=$~Lactate, Su~$=$~Succinate, FA~$=$~Focal species abundance. Shading marks $\pm 2$~SD using standard bootstrapping with 1000 sub-samplings.
\textbf{B:} CV of synthetic properties generated from a null model where all species contribute proportional to their abundance with a weight drawn from a unit uniform distribution (see text). Three synthetic properties were computed using either shuffled abundance data from Refs.~\cite{clark_design_2021,kehe_massively_2019} or by drawing random abundances from a Gaussian with mean and variance both set to 1. Null models reproduce the signature decrease of CV with richness, demonstrating that self-averaging is a generic expectation. Shading as in (A).}\label{fig:selfavg}
\end{figure}

\begin{figure*}[ht!]
\centering
\includegraphics[width=0.65\linewidth]{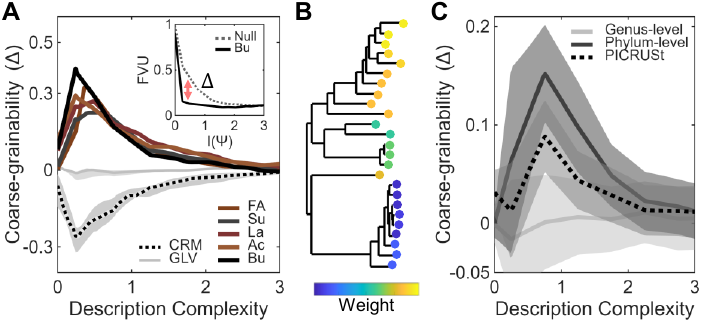}
\caption{
\textbf{Coarse-grainability arises from community structure.} \textbf{A:} Difference between the actual Pareto front and null ($\Delta$) as a function of description complexity, shown for empirical community properties measured in Refs.~\cite{clark_design_2021} and~\cite{kehe_massively_2019}, and contrasted with model-derived properties (community invadability) in two standard ecological models: random-interaction CRM and GLV. $\Delta>0$ corresponds to coarse-grainability. See inset for the actual Pareto front and null comparison in the case of Butyrate as the property of interest. For CRM and GLV, the shading represents the quartiles in $\Delta$ across 10 random invaders. For empirical curves, error bar shading is omitted for clarity but is provided in the Supplementary Text (Fig.~\ref{fig:rawCGparetos}) Abbreviations as in Fig.~\ref{fig:selfavg}. \textbf{B:} The simplest model capturing the fact that species traits are non-random: species' contributions to a synthetic property (contribution weight) are assumed to be phylogenetically correlated, with conservation depth a tunable parameter (see Methods). The phylogenetic tree corresponds to the 25 species assayed by Clark et al. (reproduced from Ref.~\cite{clark_design_2021}). Leaf coloring illustrates the case where contribution weight is deeply conserved.
\textbf{C:} $\Delta$ computed for a synthetic community property generated from the Clark et al. abundance data using phylogenetically correlated weights with conservation depth set to the genus or phylum level (see Methods). As expected, increasing conservation depth parameter leads to greater coarse-grainability. For comparison, the dashed line shows coarse-grainability for a synthetic property generated from central carbon metabolism gene profiles of the same 25 species predicted by PICRUSt2 (see Methods). Lines and shading represent median and quartiles across 10 replicate weight samplings.
\label{fig:CG}}
\end{figure*}

For all five properties measured across both datasets, we see that their coefficient of variation decreases with community richness  (Fig.~\ref{fig:selfavg}A). This observation means their value is more reproducible across high-richness communities than it is among low-richness communities, and as such, could be described as a form of ``emergent simplicity.'' However, this same behavior is observed in the simplest of null models that retain no biological structure: self-averaging is preserved even if we randomly reshuffle the measured abundances, or use a purely synthetic property to which each species contributes independently (Fig.~\ref{fig:selfavg}B; see Methods for details). This is not surprising, since both central-limit averaging and the increasing compositional similarity of high-richness samples (drawing $S>20$ species out of a pool of only 25) naturally contribute to this effect. Thus, when talking about emergent simplicity, it is essential that our terms are sufficiently well-defined to avoid any potential confusion with effects of trivial origin. To this end, from now on, we will ignore any absolute changes in the variance, assessing instead the \textit{fraction} of variance explained by various models, and examining the shape and behavior of the Pareto front in these normalized units (Fig.~\ref{fig:1}E,F). Additionally, our null-model comparisons will employ the same sampling structure as the data, and whenever appropriate, the actual measured abundance tables, so that the compositional similarity of samples is fully retained.

\subsection{Coarse-grainability arises from community structure}
We say that a community property $Y$ is coarse-grainable if coarse compositional representations are \textit{surprisingly} predictive of $Y$, compared to our null expectation. To establish the null expectation, we construct a synthetic property that uses the measured abundance table and preserves some statistics of the microscopic model fit to the actual property of interest, but assumes that species' contributions are random and uncorrelated, eliminating any notion of biological similarity or relatedness between them (for details, see Methods). By construction, the Pareto front for this synthetic property has the same endpoints as the Pareto front of interest (Fig.~\ref{fig:CG}A, inset). We quantify coarse-grainability by measuring the difference between the actual Pareto front and the null comparison, as a function of  description complexity; positive values indicate that $Y$ is coarse-grainable. Applying this methodology to the two datasets, we find that all 5 observables across both datasets show a strong signature of coarse-grainability (Fig.~\ref{fig:CG}A). Note that coarse-grainability is defined without reference to varying community richness; the analysis in Fig.~\ref{fig:CG}A is for high-richness communities ($S>20$; for Kehe et al. $S>10$).

Coarse-grainability is clearly a useful feature for understanding complex ecosystems. It is therefore worth noting that some of the dominant theoretical frameworks of high-diversity ecology are missing this property. To illustrate this, we consider two classic frameworks widely used to model high-diversity ecosystems: the Generalized Lotka-Volterra model (GLV) with a random interaction matrix, and a Consumer-Resource model (CRM) with a random resource consumption matrix, choosing ``community invadability'' as the property of interest (specifically, $Y$ is defined as the post-invasion abundance of some specified strain, initially absent from the community). This choice is convenient because it is a relevant and non-trivial community property that can be accessed in both modeling frameworks. The model parameters and simulation protocol are described in the Methods, but the outcome of the analysis is that neither model exhibits coarse-grainability (Fig.~\ref{fig:CG}A, shaded).

\begin{figure*}[t!]
\centering
\includegraphics[width=\textwidth]{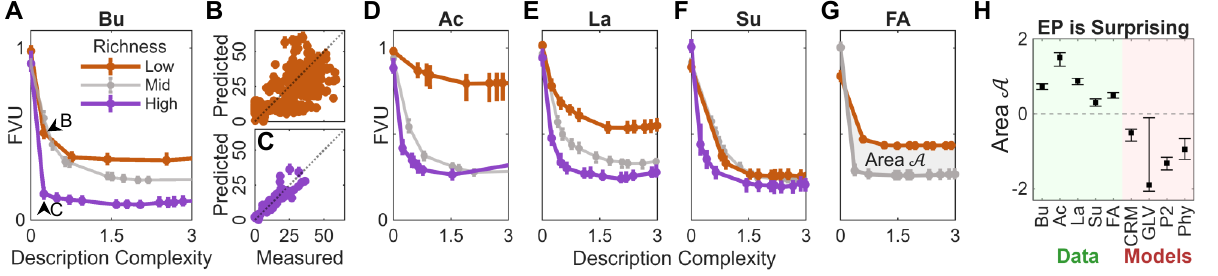}
\caption{
\textbf{Both empirical datasets exhibit emergent predictability.}
\textbf{A:} The Pareto front in error-complexity plane, computed for the butyrate measurements of Clark \textit{et al.}~\cite{clark_design_2021}. Lines show the fraction of variance left unexplained (FVU) by the best model within a given description complexity, when dataset is restricted to communities of low richness ($S
\le5$, orange), medium richness (gray) or high richness ($S>20$, purple). Points indicate the median out-of-sample FVU computed over 100 random 50-50 splits of the data into training and testing sets; error bars are the 1st and 3rd quartiles. The shift of the Pareto front indicates that at higher richness, coarse descriptions become more predictive (\textit{cf.} Fig.~\ref{fig:1}F). Panels \textbf{B-C} illustrate this result by plotting the predicted \textit{vs.} measured concentration for the two models indicated by the arrowheads. At high richness, a highly coarsened model explains 85\% of variance in butyrate production (C). In contrast, for low-richness communities, even the most predictive model of comparable complexity explains only 49\% of variance (B). \textbf{D-G:} Same as (A) for the four other properties measured in Refs.~\cite{clark_design_2021} and~\cite{kehe_massively_2019}. The Kehe~\textit{et al.} dataset (panel G) only assayed 14 isolates and is missing the high-richness tier ($S>20$ strains).
\textbf{H:} To quantify emergent predictability (EP), we measure the signed area $\mathcal A$ between the low-richness and highest-available-richness Pareto fronts of previous panels (highlighted in G). The quantification confirms that all measured properties exhibit emergent predictability. This is in contrast to all theoretical models we considered, including those that exhibit coarse-grainability (CRM, random-interaction consumer-resource model; GLV, random-interaction Lotka-Volterra; P2, PICRUSt2 central carbon metabolism; Phy, phylogenetically conserved contributions; other abbreviations as in Fig.~\ref{fig:selfavg}). Error bars show quartiles of the respective distributions. For empirical data, they represent the uncertainty of Pareto front estimation; for models, they include variability across random model instances (see Methods).
\label{fig:EP}}
\end{figure*}

The origin of this contrast is easy to pinpoint: real species and interactions are not random. If relatedness of species is taken into account, then even the simplest models can exhibit coarse-grainability. To illustrate this, we construct a simple model in which the only assumption is that closely related species contribute similarly to a community-level property. Using the abundance data from Clark et al., we define a synthetic community property to which each species contributes independently, but with weights that are correlated according to phylogenetic relatedness (Fig.~\ref{fig:CG}B shows an example). A single tunable parameter controls the phylogenetic scale at which trait correlations decay, capturing the depth of conservation (see Methods). As expected, increasing the depth of trait conservation leads to greater coarse-grainability (Fig.~\ref{fig:CG}C). Of course, the traits relevant to the measurements of Clark et al. and Kehe et al. are unlikely to have a single relevant scale of conservation depth. As an example of more realistic structure, we also include a coarse-grainability curve for a different synthetic property, constructed from the presence of carbon metabolism genes (KEGG orthologs) predicted by PICRUSt2~\cite{douglas2020picrust2} for the same set of species; see Methods for details. The shape of this curve confirms that a realistic amount of structure is sufficient to generate coarse-grainability. We note that the scale of the $y$ axis in Fig.~\ref{fig:CG}C depends on arbitrary modeling choices and is not meant to be directly compared to Fig.~\ref{fig:CG}A; our focus is on the coarse-grainability $\Delta$ becoming positive at intermediate complexity.

This analysis demonstrates that even minimal models incorporating phylogenetic signal can exhibit coarse-grainability. Some theoretical frameworks intentionally omit this relatedness structure, and thus miss this effect, but in any empirical scenario (or in models using realistic trait matrices, e.g., CRM based on measured resource preferences~\cite{ho2024resource,gralka2023genome}), at least some degree of coarse-grainability is to be expected.

\subsection{Emergent predictability is surprising and demands explanation}
Finally, we turn to emergent predictability, encoded in the behavior of the Pareto front under changes of richness. For each dataset, we resolve three tiers of community richness. We apply the same search algorithm as above to infer the respective Pareto front of each richness tier. The results are shown in Fig.~\ref{fig:EP}A,D-G.

Remarkably, we find that all five properties exhibit emergent predictability. We emphasize that the Pareto front is not an abstract concept: each point corresponds to an explicit model, and we consistently find that coarse models become increasingly predictive in communities of higher richness. As an illustration, Fig.~\ref{fig:EP}B\&C plot the measured butyrate concentration against the prediction of the simplest models identified by our framework, labeled with arrowheads in A. At high richness, a coarsening exists whereby a simple two-parameter regression explains 85\% of variance in butyrate production (Fig.~\ref{fig:EP}C). In contrast, for low-richness communities, no model of comparable complexity can explain more than 49\% of variance; the best-performing example is shown in Fig.~\ref{fig:EP}B.

As a control, this behavior is not observed if the values of $Y$ or species abundances are randomly permuted (see Supplementary Text, Figs.~\ref{fig:permuteX},~\ref{fig:permuteY}), indicating that the behavior observed in Fig.~\ref{fig:EP} is not an artifact of our analysis. The additional controls we performed include subsampling the data, perturbing the groupings forming the putative Pareto front, and considering alternative metrics of complexity $I(\Psi)$; see Supplementary Text, Figs.~\ref{fig:down_sample},~\ref{fig:relabel_test},~\ref{fig:num_classes}, respectively. We conclude that emergent predictability is robustly supported by data.

Crucially, however, none of the standard models show this behavior. To see this, we quantify emergent predictability by measuring the area contained between the high-richness and low-richness Pareto fronts (see Methods).  Fig.~\ref{fig:EP}H summarizes the contrast between the empirical results and all the theoretical models discussed so far. Each of these models exhibits self-averaging, and multiple models exhibit some degree of coarse-grainability, provided they retain some notion of relatedness between species (PICRUSt; phylogenetically conserved contributions). However, none of these models exhibit emergent predictability. (See also Supplementary Text section~\ref{si:noEP_from_poolStruct}.)

\subsection{Mechanistic origin of emergent predictability}
\begin{figure}[b!]
\centering
\includegraphics[width=0.95\linewidth]{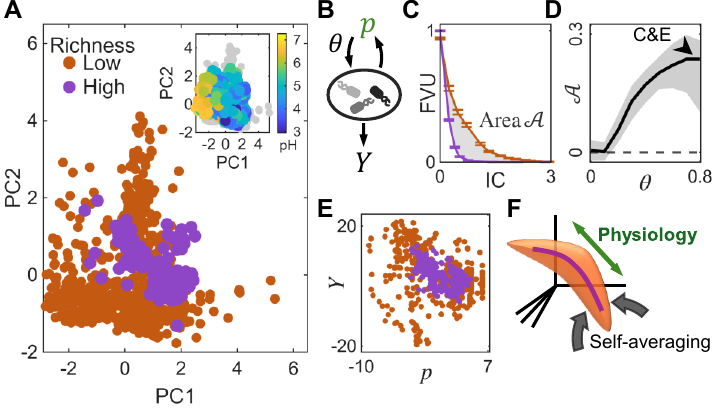}
\caption{
\textbf{Mechanistic origin of emergent predictability.} \textbf{A:} Principle component analysis (PCA) of the four-dimensional functional measurements of Clark \textit{et al.} (butyrate, acetate, lactate, succinate) indicates that high-richness samples form a lower-dimensional subspace. Coloring the PCA plot by measured pH (inset) reveals that the dominant axis of functional variation is correlated with community pH. Principal axes computed for the dataset as a whole; mid-richness communities are omitted for clarity (but included in the inset). \textbf{B:} A minimal model that can reproduce emergent predictability. The key ingredient is a feedback loop (of strength $\theta$) with a shared environmental variable $p$. Species differ by their preferred interval of $p$ which allows them to be active; see text for details. Species' contribution strengths to the property of interest $Y$ are random and uncorrelated (no special structure is assumed). \textbf{C:} The model exhibits emergent predictability; shown is an example for feedback strength $\theta=0.8$. Richness tiers match Clark \textit{et al.}, with 800 simulated communities per tier (see Methods). \textbf{D:} In the model, emergent predictability is controlled by feedback strength $\theta$. As in Fig.~\ref{fig:EP}, emergent predictability is quantified as the signed area between low- \& high-richness Pareto fronts (shaded in panel C). Error bar shading shows quartiles over 10 random model instances. The feedback strength corresponding to the example shown in panels C \& E is highlighted. \textbf{E:} Scatter plot of the community property $Y$ versus the environmental variable $p$ for the same simulated dataset as used in panel C; colors as in (A). The plot offers an intuitive explanation of the origin of emergent predictability in the model. At higher richness (purple), the point cloud is tighter, but the tightening is anisotropic, allowing $p$ (and, by extension, $p$-dependent compositional features) to become more predictive of $Y$.
\textbf{F:} A cartoon graphical summary of the proposed general mechanism illustrated by our model. Communities are represented as points in an abstract ``functional space.'' As richness increases, communities generally become more functionally similar due to generic forms of self-averaging. However, if physiological or environmental feedbacks oppose self-averaging along some axes of community variation (e.g., the $p$ axis in our model), then at high richness, such axes will become responsible for an increasingly larger fraction of variance in related properties.
\label{fig:pH}}
\end{figure}

At least for the system of Clark \textit{et al.}, the emergent predictability appears to be related to the dominant structuring role of pH, which is one-dimensional. While the data is insufficient to demonstrate this directly (independent measurements of pH were conducted for only a subset of communities), we can obtain strong indirect evidence by leveraging the fact that four functional properties were measured simultaneously. These four measurements (the final concentrations of butyrate, acetate, succinate and lactate produced) define, for each community, a point in what we will call the ``functional space''. Fig.~\ref{fig:pH}A shows that at high community richness, these points are found along a lower-dimensional subspace. Indeed, for high-richness communities, 73\% of variance is explained by PC1, compared to only 41\% for low-richness communities (Fig.~\ref{fig:si_fig:pca}). This dominant axis of variation is strongly correlated with community pH (Fig.~\ref{fig:pH}A, inset), consistent with  Ref.~\cite{clark_design_2021} identifying pH-mediated interactions as a key factor structuring these communities. This observation likely explains the sudden drop in the Pareto fronts observed in Fig.~\ref{fig:EP}A,D-G. Indeed, any combination of species whose abundance is correlated with pH will explain a substantial fraction of variance in any of the community properties that have a loading on the dominant principal component of Fig.~\ref{fig:pH}. As a result, even a little compositional information can be sufficient to improve predictive power, which is what the Pareto front indicates. Crucially, the analysis of Fig.~\ref{fig:EP} used no knowledge of pH. Thus, our framework may provide a prior-knowledge-agnostic way for discovering the existence of strong ecosystem-structuring factors.

For specifically Fig.~\ref{fig:EP}A (butyrate), the results of Clark \textit{et al.}~\cite{clark_design_2021} provide some additional mechanistic insight. Of the 25 strains in this study, five are capable of producing butyrate. However, in high-richness communities, butyrate production is dominated by one of them, \textit{Anaerostipes caccae} (AC). And indeed, the first point on the high-richness Pareto front in Fig.~\ref{fig:EP}A is the grouping \{AC, all other strains\}, and provides an excellent prediction for the amount of butyrate produced (Fig.~\ref{fig:EP}C). In contrast, in low-richness mixtures, all five strains contribute to the production, and no single compositional feature is similarly predictive.

Clark \textit{et~al.} hypothesized that this transition is associated with AC shifting at high community richness to an alternative metabolic pathway, utilizing lactate rather than sugars, a physiological switch mediated by sensing pH and/or the depletion of key resources. Identifying candidate mechanistic explanations for the other properties, and validating these hypotheses, would require new experiments. It is conceivable that the observations of Fig.~\ref{fig:EP}A, D-G are all a consequence of some idiosyncrasies of specific taxa. However, the common outcome observed across two independent datasets suggests that emergent predictability as quantified here may have a more general origin.

\subsection{Candidate general mechanism: ecological feedback that opposes self-averaging}
We will now describe and support a candidate general mechanism for emergent predictability to arise. Somewhat counter-intuitively, we propose that emergent predictability arises when some physiological or environmental feedback \textit{opposes} statistical self-averaging along certain axes of community variation. To illustrate, we construct a simple model (Fig.~\ref{fig:pH}B) inspired by Refs.~\cite{ratzke2018modifying,amor2020transient,narla2025dynamic}. Let species $i$ be characterized by metabolic activity $a_i$. The metabolic activity defines its growth, scales its contribution to a collective property of interest $Y$, but also its influence on a shared environmental variable $p$ (e.g., pH, osmolarity, oxygen content, toxin concentration):
\begin{align*}
    \dot{n}_i &= n_i a_i&&\text{population growth}\\
    \dot{r} &= -\sum_i a_i n_i &&\text{nutrient depletion}\\
    \dot{p} &= \sum_i a_i n_i c_i &&\text{dynamics of $p$}\\
    \dot{Y} &= \sum_i a_i n_i y_i &&\text{dynamics of $Y$}.
\end{align*}
We will assume no special structure in species' contributions to either $p$ or $Y$: both $c_i$ and $y_i$ will be drawn randomly and independently from Gaussian distributions. However, crucially, we will assume that the activities $a_i$ are impacted by feedback from $p$. Specifically, each species $i$ has a different interval of $p$ that it prefers, and can only be active within a finite range of a preferred value (denoted $\rho_i$):
\[a_i = r~F(|p - \rho_i|).\]
Here, $r$ is nutrient availability, and $F$ has a finite width $\simeq \Delta p$ (we pick $F$ to be a truncated Gaussian; see Methods). As an example, a species acidifying its own environment can poison its own growth in monoculture, even if nutrient is still available~\cite{ratzke2018ecological}. The parameter $\theta=1/\Delta p$ encodes the feedback strength: large $\theta$ corresponds to species highly sensitive to changes in $p$ (a narrow activity range), whereas $\theta=0$ corresponds to $p$ being irrelevant. This feedback, assumed to be mediated by species physiology, is the key ingredient of our model. For more details, see Methods.

Fig.~\ref{fig:pH}C confirms that this model displays emergent predictability (simulation parameters are listed in Methods). As we expected, the magnitude of the effect is controlled by feedback strength $\theta$ (Fig.~\ref{fig:pH}D). Intuitively, this can be understood as follows. In the analytically solvable limit $\theta\rightarrow 0$ (no feedback), it is easy to see that $p$ is irrelevant, the species' contributions to $Y$ are subject to simple self-averaging, and there is no emergent predictability (for more discussion, see Supplementary Text, section~\ref{si:noEP_from_poolStruct}). At non-zero $\theta$, self-averaging continues to play a role leveling out the quirks of individual species within a tier of similar $p$ preference. However, along the $p$ axis, the $p$-mediated ecological feedback opposes self-averaging. For instance, a species that is strong driver of $p$ (large $c_i$) will always drive this environmental variable to the upper limit of its viable range, which is a fixed value set by species physiology ($\simeq \rho_i+\Delta p$) and does not depend on richness. As a result, the axis of variability associated with $p$ preference maintains its relevance while other differences self-average, and so becomes responsible for an increasingly larger fraction of the total variance of the property of interest (illustrated in Fig.~\ref{fig:pH}E,~F). A more detailed analysis of this mechanism can be found in the Supplementary Text.

The feedback of the kind we described is expected to be widespread~\cite{narla2025dynamic}, and Fig.~\ref{fig:pH}D demonstrates it is sufficient to endow a system with emergent predictability. Thus, our analysis suggests that emergent predictability is a nontrivial and exciting feature that might well be ubiquitous in natural communities.

\section{Discussion}
The promise of emergent simplicity is to enable predictions that work because of diversity, not despite it. However, so far, the field was unable to reach a consensus on whether emergent simplicity is real, whether it is surprising, and whether it is useful. Answering the first question requires a quantitative definition; the second requires a null model for comparison. For the third, we need an explicit focus on prediction rather than reproducibility, and a clear understanding of whether emergent simplicity is a biological statement, or a statistical one. Indeed, predictions that are a direct mathematical consequence of the number of species being large, independent of the underlying biology, are unlikely to be sufficiently nontrivial to be useful in practice.

Here, we developed an information-theoretic framework to quantify the notion of emergent simplicity, and applied this framework to data from two separate experiments on microbial communities assembled in the lab. Our approach enabled us to define and distinguish coarse-grainability and emergent predictability, and to contrast both of these phenomena with ``emergent reproducibility'' (self-averaging) considered in much previous work.

All empirical properties we considered showed a clear signature of coarse-grainability. However, as we discussed, this is not necessarily surprising since for any real ecosystem, at least some degree of coarse-grainability is expected. This does not mean that coarse-grainability is uninteresting. Some empirical reports indicate a truly striking extent of coarse-grainability~\cite{jeglot_microbiome_2021}, and it is not clear that this can be fully explained by the relatedness of community members, especially given the extensive,  functionally relevant diversity that is known to persist even at the level of strains~\cite{goyal_interactions_2021}. A clear quantitative metric of coarse-grainability, as presented here, should help investigate this in future work.

Our main findings concerned emergent predictability, whereby coarser compositional descriptions become more predictive at higher richness. Our key result is that this phenomenon is robustly supported by empirical data; yet is absent from standard models of theoretical ecology, making it highly non-trivial and surprising.

In four of the five cases (measurements of Clark \textit{et al}.), the phenomenon appeared to be linked to the dominant structuring role of pH in these communities. This observation echoes other recent examples where coarse-grained predictions are enabled by general physiological~\cite{estrela_functional_2022}, metabolic~\cite{vila2023metabolic}, or toxicity-mediated constraints~\cite{crocker2023Environmentally}. We described a mechanism explaining how structuring factors of this kind may acquire a higher explanatory power at higher diversity. We illustrated it in a minimal model demonstrating how emergent predictability can arise from a physiology-mediated feedback with a collective environmental variable.

Perhaps the most exciting implication of our analysis is that it reconciles the generality of the central limit theorem with the possibility of nontrivial predictions. Indeed, under the mechanism we proposed, the diversity-dependent simplification is ultimately still a form of self-averaging -- but one that leads to a stronger collapse onto a lower-dimensional subspace governed by metabolic or physiological constraints (Fig.~\ref{fig:pH}F). Understanding this behavior will require extending the theoretical investigations of large-$N$ ecosystems beyond the Gaussian-interaction regimes~\cite{cui2021diverse,marsland_iii_available_2019,azaele2024generalized}, and integrating these ongoing efforts with new models capable of capturing the role of such structuring factors.

Our approach has limitations. Our definitions of emergent simplicity required specifying a class of models, and here, due to the limited size of the available datasets, our choice (linear regression) was particularly simple. However, our theoretical framework is more general; with sufficient data, the natural generalization of the Pareto fronts in Figs.~1, 3-5 would be to evaluate the predictive power of machine learning models as a function of the dimension of the compressed (latent-space) representation of community composition, similar to the approach of Ref.~\cite{plata2025designing}. Understanding the extent to which emergent predictability depends on the choice of model class remains a question for future work. We should also note that in general, community richness (number of species) is a poor metric of community diversity, as it is overly sensitive to detection thresholds, and fails to account for the relatedness of community members~\cite{peet1974measurement}. This was not an issue for our work, as the synthetic communities considered here were assembled from a small pool of well-delimited taxa. However, for natural communities, this simple metric of diversity would need to be replaced by one of the better-suited metrics developed in the ecological literature~\cite{iknayan2014detecting,mendes2008unified,crupi2019measures}.

All simulations performed in \textsc{MATLAB} (Mathworks, Inc.). The associated code, data and scripts to reproduce all figures in this work are publicly available~\cite{MENDELEY_REPO}.

\begin{acknowledgements}
We thank S.~Allesina, A.~Goyal. K.~Gowda, T.~GrandPre, C.~Holmes, S.~Kaplan, S.~Kuehn, R.~McGee, A.~Murugan, O.~Venturelli and K.~Wood for helpful discussions. MT acknowledges a CAREER award from the National Science Foundation (PHY-2340791). This work was also supported by the National Science Foundation grant PHY-2310746, and the Michigan Pioneer Fellows Program award to JM.
\end{acknowledgements}

\bibliographystyle{abbrv}
\bibliography{biblio}

\cleardoublepage

\section*{Materials and Methods}\label{main:methods}

\subsection*{Data selection and processing}
Our criteria of data selection, and how they apply to available studies, are presented in Sup. Table S1. These criteria selected two datasets for our analysis, Ref.~\cite{clark_design_2021,kehe_massively_2019}. The raw data we used is published alongside the original studies, and/or is available on Mendeley Data~\cite{MENDELEY_REPO}. Data from Kehe~\textit{et~al.}\ were generated using microwell arrays for random community assembly using nanoliter droplets of defined soil isolates, minimal media, and sucrose. Community culturing experiments of Clark~\textit{et~al.}\ were performed in an anaerobic chamber with defined isolates assembled in a chemically defined medium to support the growth of diverse gut species. Further experimental details and data preprocessing steps are described in the corresponding publications~\cite{clark_design_2021,kehe_massively_2019}.

For both  datasets, we resolve three tiers of community richness (with the high-richness tier, $S>20$ strains, only available for Clark \textit{et~al.}~\cite{clark_design_2021}). Since Clark \textit{et~al.}\ did not directly measure absolute strain abundances in a community (only its total OD and the compositional makeup assessed by 16S sequencing), we adopted an approach where the relative abundances estimated from the 16S data were first scaled by community OD (of a given sample), and then standardized to zero mean and unit variance (across samples) before training any regression models. For consistency, the same standardization procedure was also applied to Kehe \textit{et~al.} data; though omitting this step does not qualitatively change the outcome in either case; see Supplementary Text, section~\ref{si:norm_predictors}, and Fig.~\ref{fig:non_standardized} and for further discussion.

\subsection*{Model-derived synthetic community properties}
When using simulated data, our analysis employed models of two kinds: (1) Effective models that adopted a simple linear-contribution ansatz to map community composition directly into a synthetic ``community property,'' and (2) mechanistically-informed dynamical models.
\subsubsection*{Effective models}
These models were employed as a simple setting to establish the \textit{minimal} conditions required for the emergence of each of the flavors of emergent simplicity we defined. Accordingly, in all cases, the synthetic property $\tilde Y_\mu$ was defined by a simple linear-contributions model:
\begin{equation}
\tilde{Y}_\mu \equiv \sum_i \tilde{\beta}_i \hat{n}_{i \mu} + \epsilon_\mu.\label{eq:linearAnsatz}
\end{equation}
Here, $\hat{n}$ denotes standardized abundances (unless specified otherwise, we used the abundance table measured in Clark \textit{et. al.}). The term $\epsilon_\mu$ is Gaussian random noise with mean 0; its variance directly sets the FVU of the best predictive model. Finally, the weights $\tilde{\beta}_i$ determine the strength of species' contributions to this synthetic property $\tilde Y$. By construction, with sufficient data, the best microscopic model fit to such a synthetic dataset would be the model with coefficients $\beta_i$. The different models discussed below differ primarily by the choice of these ground-truth $\tilde \beta_i$ (random and uncorrelated; correlated with phylogenetic relationships; or derived from a realistic trait table constructed using PICRUSt2).

\paragraph{Null Model for Self-averaging:}
For Fig.~\ref{fig:selfavg}B analysis, we construct the simplest null model by drawing $\tilde{\beta}_i$ in Eq.~\eqref{eq:linearAnsatz} from a unit uniform distribution. For species abundance inputs we use either randomly shuffled abundance data from Clark~\textit{et~al.} or Kehe~\textit{et~al.}, or random values drawn from a Gaussian with mean and variance both set to 1; the three choices correspond to the three lines shown in Fig.~\ref{fig:selfavg}B. Note that in Fig.~\ref{fig:selfavg}B, our goal is to show that self-averaging can arise with no biological structure whatsoever, which is why we \textit{shuffle} the empirical abundances (or just use Gaussian-random values instead). Specifically for the simulations shown in Fig.~\ref{fig:selfavg}B, the $\epsilon_\mu$ term was set to zero.

\paragraph{Null Model for Quantifying Coarse-grainability:}
In order to quantify whether the Pareto front constructed for some property $Y$ descends ``faster than expected'', we generate a synthetic community property $\tilde{Y}$ whose Pareto front can serve as a null-model comparison. To make the comparison conservative, $\tilde Y$ should be constructed based on the same abundance table (no shuffling!). To make the comparison fair, $\tilde Y$ must satisfy two constraints: its Pareto front should start at FVU~$\approx 1$, as is the case for all our empirical properties (``constraint~1''); and the FVU attained by the most predictive model should match that of the reference property $Y$ (``constraint~2'').

To this end, the weights $\tilde{\beta}_i$ in Eq.~\eqref{eq:linearAnsatz} were drawn from a normal distribution with zero mean, and variance matching the spread in the regression coefficients $\beta_i$ of the ``microscopic'' model constructed for the reference property $Y$. Centering $\tilde\beta_i$ at zero ensures that the biomass-only description has no predictive power, satisfying constraint 1. Then, the variance of the added Gaussian noise $\epsilon_\mu$ was adjusted for $\tilde Y$ to satisfy constraint 2. (Practically speaking, we adjusted its variance so that the \textit{third} lowest point of the Pareto front of $\tilde Y$ matches the third lowest point of the Pareto front of $Y$. This ensures robustness to the stochasticity inherent to estimating out-of-sample FVU by cross-validation.)

This protocol ensures that the two Pareto fronts (for $Y$ and $\tilde Y$) have matching endpoints, and allows us to ask whether the speed of the Pareto front descent observed for $Y$ (its ``coarse-grainability'') is noteworthy, or is no faster than can be reproduced by the simplest null model described.

\paragraph{Phylogenetically Conserved Contributions:}
In Fig.~\ref{fig:CG}B,~C, we sought to demonstrate that even the simplest models implementing the intuition that ``some species are more similar than others'' can exhibit coarse-grainability. For this, rather than choosing $\tilde{\beta}_i$ in Eq.~\eqref{eq:linearAnsatz} to be independent and uncorrelated, we drew them in a phylogenetically-informed way. Specifically, we first used the phylogenetic tree of Clark \textit{et al.} species (as presented in the original reference Ref.~\cite{clark_design_2021}) to compute the pairwise patristic distance matrix $D_{ij}$, normalized to set the largest distance to 1. Then, the contribution weights $\tilde{\beta}_i$ were drawn from a multivariate Gaussian distribution with zero mean and the correlation matrix
\[C_{ij} = \frac{1}{1 + \mathrm{exp}(\Delta_d (D_{ij} - d))}.\]
This sigmoid functional form was chosen to implement the notion of a finite conservation depth: the contributions $\tilde \beta_i$ of closely related species are highly correlated, but this correlation falls off sharply beyond a distance threshold $d$ (the steepness parameter is set to a large value $\Delta_d=50$). The standardized abundance matrix $\hat{n}_{i\mu}$ we used in Eq.~\eqref{eq:linearAnsatz} was based on Clark \textit{et al.} measured abundances, combined across all richness tiers to have the most data (as a reminder, coarse-grainability is defined without reference to changing richness). The variance of $\epsilon_\mu$ was adjusted to ensure that the best-performing model on the Pareto front explained 70\% of simulated data (matching the typical value attained by the empirical Pareto fronts at high richness).

The distance threshold $d$ is the key parameter varied in the analysis of Fig.~\ref{fig:CG}C. By computing the typical distance between pairs of leaf nodes (Clark \textit{et al.} species) that fall within a given taxonomic level, we can obtain values for the depth parameter $d$ that correspond to genus- or phylum-level conservation of trait correlations. This allowed us to label the curves of Fig.~\ref{fig:CG}C in an interpretable way (e.g., ``phylum based conservation''), rather than by the corresponding value of $d$.

\paragraph{Realistic contribution structure (PICRUSt-based  model):}
Finally, we wanted to consider an example with a more realistic structure. To this end, we constructed another model where instead of drawing the weights $\tilde\beta_i$ from some distribution, we derive them from an actual matrix of predicted organism traits.

Specifically, we first generated a phylogenetic tree based on the 16S sequences in Clark \textit{et al.} using SEPP~\cite{mirarab2012sepp}. Next, we applied PICRUSt2~\cite{douglas2020picrust2} to generate a predicted presence/absence table of KEGG orthologs (KOs) belonging to the Carbohydrate Metabolism category (chosen as one general category relevant for the fermentation end-products assayed in the Clark \textit{et al.} study). Any KOs that were present in no species, or in all of them, were discarded. This procedure mapped each species into a 123-dimensional binary ``KO profile'' $\vec P_i$. Then, for each species $i$ we computed the dot product $w_i$ of its KO profile with a fixed vector $\vec v$: $w_i=\vec P_i\cdot \vec V$. For simplicity, we picked $\vec V=(1, 1, \dots 1)$. While this mapping is, of course, still simplistic, it inherits some of the complexity of the PICRUSt-generated trait table, and reflects the fact that relevant organism traits do not typically possess a single scale of conservation depth. Finally, the weights $\tilde\beta_i$ were constructed by shifting the distribution of $w_i$ to have zero mean: $\tilde\beta_i=w_i-\langle w_i\rangle$. Without this shift, the community biomass alone would already be highly predictive of the property, which would mask the effect we seek to describe. (See important additional discussion in the Supplementary Text Section~\ref{si:norm_response}).

\subsubsection*{Dynamical models}
Beyond the effective models described above, our analysis also used three mechanistically-informed dynamical models: two representatives of standard modeling frameworks (GLV and CRM) with ``community invadability'' as the property of interest, used in Fig.~\ref{fig:CG}\&\ref{fig:EP}; and a physiological feedback model used in Fig.~\ref{fig:pH}.

\paragraph{Generalized Lotka-Volterra model with random interactions:}
The GLV model we consider takes the following form:
\[
\dot{n}_i = \frac{r_i}{K_i} n_i \left( K_i - n_i - \sum_{j,j\neq i} a_{ij} n_j \right),
\]
which describes the abundance dynamics of each strain $i$ in terms of its growth rate $r_i$, carrying capacity $K_i$ and interactions with other strains encoded in $a_{ij}$. For simplicity, we set the growth rates for all strains to be the same $r_i=1$ since the key parameters determining equilibrium properties are the $K_i$ and interactions $a_{ij}$~\cite{barbier_generic_2018}. We then adopt the approach common in large-$N$ theoretical ecology, drawing strain interactions $a_{ij}$ and carrying capacities $K_i$ randomly from specified distributions. Following refs.~\cite{bunin_interaction_2016,barbier_generic_2018}, we use a normal distribution for both $K_i$ and $a_{ij}$. For the $K_i$, we set the mean to 1 and $\mathrm{std}(K_i) = 0.3$. For the interactions $a_{ij}$, the mean and spread of the distribution scale with the number of strains as $\mu / N$ and $\sigma / \sqrt{N}$, respectively. We set $\mu = 1$ and $\sigma = 0.8$ (modified from the ``simplest'' choice $\mu = 1$ and $\sigma = 1$ to facilitate the construction of a coexisting species pool). To obtain a strain library from which to subsample as illustrated in Fig.~\ref{fig:1}A, we first simulate the dynamics of a larger pool to find a set of coexisting strains (abundance $n_i > 10^{-6}$). With our parameter choices, we found that a random set of 45 strains resulted in a community of 26 coexisting strains, similar to the high richness communities of the empirical datasets we consider in this work.

\paragraph{Consumer-Resource model with a random consumption matrix:}
The CRM model describes the abundance dynamics of strains competing for a set of resources. We adopt the following parameterization (identical to Ref.~\cite{moran_defining_2022} with slightly adjusted notation):
\begin{align*}
    \dot{n}_i &= n_i \left( \sum_\alpha \sigma_{i\alpha} h_\alpha - \chi_i \right) \\
    h_\alpha &= h_\alpha(T_\alpha) \equiv \frac{b_\alpha}{1+T_\alpha/K_\alpha} \\
    T_\alpha &\equiv \sum_i n_i \sigma_{i\alpha}.
\end{align*}
Briefly, each strain $i$ is described as a binary phenotype $\sigma_{\alpha i}\in\{0, 1\}$ indicating which resources $\alpha$ each strain can or cannot consume. The growth benefit of each resource $h_\alpha$ is depleted as the total demand $T_\alpha$ for resources increases with increasing abundance of competing strains. The strength of this depletion is set by parameters $b_\alpha$ and $K_\alpha$. For simplicity, we take $b_\alpha = 1$ for all resources $\alpha$. Finally, the relative competitiveness of strains is reflected in their ``maintenance cost'' $\chi_i$; we follow Ref.~\cite{tikhonov_collective_2017} to adopt $\chi_i = c + \sum_\alpha \lambda \sigma_{i\alpha} + \delta_i$. As in ref.~\cite{moran_defining_2022}, we set the baseline cost $c=0.1$ and the cost per resource consumed $\lambda = 0.5$. Using the same approach as in studying the GLV model, we draw parameters $K_\alpha$, $\delta_i$ and $\sigma_{i\alpha}$ randomly. The ``carrying capacities'' $K_\alpha$ are Gaussian distributed with mean $10^{10}$ and standard deviation $10^9$. (Note that the scale of abundance is of course arbitrary, and could have been set to 1. The value of $10^{10}$ is chosen to mimic the large typical size of microbial populations.) Random contributions to each strain's cost $\delta_i$ are also Gaussian-distributed with mean 0 and standard deviation of 0.01. Each strain phenotype $\vec{\sigma}_i$ is a random $L$-dimensional binary vector where each element represents the ability to consume a resource, and has probability $p=0.5$ to be 1. Just as we did in the GLV model above, we assemble a community from a larger pool of $M$ phenotypes, choosing $M=50$ and $L=40$ such that the number of coexisting strains is similar to the empirical datasets ($\simeq$25).

\paragraph{Simulated invasion experiments: }
In both the GLV and CR models, we use the same protocol for our computational experiments, which follows the scheme outlined in Fig.~\ref{fig:1}A. Using the same distributions used to form the complete community, we randomly draw a focal strain to invade subsets of this community. Specifically, we assemble 1000 communities for each richness tier: low richness with 1 to 5 strains, mid richness with 11 to 15 strains, and high richness with 21 to 25 strains. After a specified amount of simulation time, we note the abundances of all community members, recording them as the ``pre-invasion abundances''; these will be used as predictors. We then computationally ``invade'' the communities with the focal strain, and measure the post-invasion abundance of the focal strain after the same amount of simulation time has elapsed; this abundance is recorded as the response variable we seek to predict.

\paragraph{Physiological feedback model: }
We take as inspiration the model presented in Refs.~\cite{ratzke2018modifying,amor2020transient,narla2025dynamic} of pH-mediated interactions between microbes, and we generalize it to a consumer-resource context. Each species $i$ is characterized by its strength $c_i$ in modifying an environmental variable $p$ and by its preferred environmental variable $\rho_i$, which in nature would be set by physiological constraints. Species in a community interact by competing for a common nutrient source $r$ and by modifying the environmental variable. The growth rate of species $i$ is expressed as
\[a_i = r~\mathrm{max}(f(p - \rho_i)-\epsilon,0),\]
where $f(p-\rho_i)\equiv e^{-\frac{(p-\rho_i)^2}{2 \sigma^2}}$ sets the range $\sigma$ of growth around the preferred environmental pH and $\epsilon$ sets a viability threshold outside the preferred environmental variable. The resulting behavior is that as the environmental variable moves further away from a species' preferred state, that species begins to consume and grow less, until eventually it stops consuming and growing entirely. The abundance and environment dynamics are expressed as follows:

\begin{align*}
    \dot{n}_i &= a_i n_i\\
    \dot{r} &= -\sum_i a_i n_i = -\sum_i \dot{n}_i\\
    \dot{p} &= \sum_i a_i n_i c_i = \sum_i c_i \dot{n}_i
\end{align*}

Note that this is a batch culture model, rather than a steady-state chemostat model. Using a batch culture model allows us to meaningfully tune community richness with all initial species contributing to the dynamics, in contrast to a chemostat version which would only permit at most 2 species surviving at steady state, regardless of their initial number.

Finally, to incorporate some community-level property $Y$ we wish to predict using the final species abundances, species are assigned a weight $y_i$ describing the effect of their metabolism on said property. We chose the dynamics of property $Y$ to be similar to the environmental variable:
\[\dot{Y}=\sum_i a_i n_i y_i = \sum_i y_i \dot{n}_i.\]
The simulations of Fig.~\ref{fig:pH} used the following parameters: $c_i$ and $y_i$ both drawn independently from a Gaussian with mean 0 and unit variance; $\rho_i$ drawn from a Gaussian with mean 0 and standard deviation 5; and $\epsilon$ set to 0.0111, equivalent to species arresting growth when the environmental variable is beyond $\pm$3 of their preference when feedback strength is 1. Each species was initialized with 0.05 abundance, the initial value of $r$ was fixed at 10. A fixed pool of 25 species was used in each simulation, with richness tiers defined as for all other models and empirical datasets, and 800 unique random communities assembled per richness tier. Once the simulation reaches a steady-state (when the community resource is depleted, or the environmental variable too extreme for any species to grow), the steady state abundances are used as predictors of the community property of interest.

\subsection*{Constructing \& manipulating the Pareto fronts}
\subsubsection*{The Metropolis-Hasting algorithm}
The Pareto front of coarsenings $\Psi$ consists of those with minimal prediction error for a given complexity $I(\Psi)$. An exhaustive search through the space of all possible coarse-grainings is computationally infeasible unless the library of strains is very small: the total number of ways to partition $S$ strains (the Bell number $B_S$) grows (super-)exponentially with S, and for $S=25$ taxa as used in the Clark \textit{et~al.} study, $B_{25}\approx 5\times10^{18}$. This necessitates developing algorithms to approximate the Pareto front via a heuristic search.

Here, we leverage the approach recently described in~\cite{zhao2024linear}, adopting a variant of the Metropolis-Hastings algorithm to search the space of possible coarsenings and estimate the Pareto front by retaining the most predictive ones. Briefly, the algorithm initializes a random initial guess for the Pareto front by drawing a random set of candidate coarsened groupings and evaluating the description complexity and prediction error of each grouping. Then, the algorithm main loop: (1) constructs a new candidate grouping by randomly splitting or merging a randomly selected grouping from the current set, (2) evaluates the information-prediction attributes of the new candidate, and (3) if the new grouping improves on the prediction error of the model currently memorized in the respective complexity bin, the memorized model is overwritten by the new candidate. Otherwise, the replacement is probabilistic with a Boltzmann probability set by the difference in their prediction errors. The main loop is repeated $10^5$ times for each of 15 random initializations, and the top-performing groupings are aggregated into the final ``best guess'' of the Pareto front models. The performance of these best-guess groupings are then cross-validated to obtain a distribution of out-of-sample FVU measurements (see Supplementary Text, section~\ref{si:cv_of_lin_reg}). Finally, we perform the following convergence check: (a) for each coarsening $\Psi$ making up the best Pareto guess, generate all neighboring coarsenings that successively re-assign each single strain to all other classes in $\Psi$, (b) identify and include any relabeled coarsenings that result in a median FVU that fall below the convex hull spanned by the lower quartiles in FVU of the current Pareto, (c) repeat steps (a) and (b) until no single-strain relabelings of the Pareto front coarsened descriptions fall below the lower FVU quartiles. At this point, the algorithm has converged to at least a local optimum. The output returned by the algorithm consists of a set of coarsenings $\{\Psi\}$, and for each, 100 estimates of out-of-sample FVU, computed by cross-validation over 50-50 splits. The error bars shown in panels~\ref{fig:EP}A,D-G and \ref{fig:pH}C are the quartiles of these respective distributions.

Additional details on our methodology of constructing and testing the Pareto fronts for robustness and artifacts, and the actual computed Pareto fronts used for all analyses in the main text can be found in the Supplementary Text section~\ref{si:pareto_details}.

\subsubsection*{Aggregating the Pareto fronts}
When we consider synthetic properties where species contributions $\tilde{\beta}_i$ are drawn from a distribution (i.e., the coarse-grainability null model, PICRUSt2 model, and phylogenetic-based model), each random sample of these weights defines its own instance of the synthetic property $\tilde{Y}$, and thus a distinct Pareto front. To summarize this ensemble, we  aggregate it into a single Pareto front (with larger error bars), using the following procedure.

As mentioned above (and detailed in Supplementary Text, Sec.~\ref{si:cv_of_lin_reg}), each Pareto front consists of a set of coarsened models identified by the algorithm, and for each such model (a grouping $\Psi$), its performance is reported as a set of 100 FVU values computed via cross-validation over 100 folds of splitting the data into training and testing sets. When aggregating an ensemble of Pareto fronts, these cross-validation sets are collated together and binned along the axis of description complexity $I(\Psi)$. We then compute the median FVU within each complexity bin; this is the aggregated Pareto front, which is now defined by its vertices at bin center locations, but which we interpolate into a piecewise linear curve. To report an error bar, we similarly aggregate and compute a running average of the magnitude of deviations (relative to this \textit{interpolated} aggregated curve), and report the quartiles of the distribution of these deviations.

\subsubsection*{Comparing the Pareto fronts}
When assessing coarse-grainability (Fig.~\ref{fig:CG}), our task is to compare an actual Pareto front (constructed for some property $Y$, either empirical or model-derived), and the (ensemble of) null-model comparisons $\tilde Y$. To perform this comparison, we first aggregate this ensemble into a single null-model Pareto front as described above. Now, we have two Pareto fronts (with error bars), and our task is to construct a difference curve (also with error bars). To do so, we first interpolate both the Pareto fronts and the error bars, so they are defined at same values of the complexity axis. Then, the difference is trivially computed by subtracting the interpolated curves. As for the error bars, the uncertainties in estimating both curves are independent, and we combine them in quadrature: $\sigma=\sqrt{\sigma_1^2+\sigma_2^2}$.

\subsubsection*{Area between the Pareto fronts}
In Fig.~\ref{fig:EP}H, we quantify emergent predictability by computing the area between the high- and low-richness Pareto fronts. In order to compute the uncertainty of its estimation, we compute 100 areas between 100 noisy samples of each of the Pareto fronts (sampling from the 100 FVU values, reported for each model), and record the median and the quartiles of this distribution. Fig.~\ref{fig:pH}D used the same approach, but the areas were additionally aggregated over 10 random instances of the model.

It is important to emphasize that this area-based metric is purely heuristic (note that the area is measured on a log plot, with an arbitrary x-axis cutoff of $3$ bits of entropy). This approach is helpful for presentation purposes, affording the simplicity of summarizing several comparisons in a single panel (Fig.~\ref{fig:EP}H). The most direct evidence is to compare the full profiles (as we do for coarse-grainability in Fig.~\ref{fig:CG}); this analysis is presented in Supplementary Text, section~\ref{si:raw_paretos}.

\newpage
\onecolumngrid
 \section*{Supplementary Text}
 \appendix
 \setcounter{figure}{0}
 \setcounter{equation}{0}
 \renewcommand{\theequation}{S\arabic{equation}}
 \renewcommand{\thefigure}{S\arabic{figure}}
 \renewcommand{\thetable}{S\arabic{table}}

\section{Data selection: Other empirical datasets available}
Scanning the literature for empirical studies that assemble communities from a fixed library of strains, we identified 9 candidates for further consideration. These are listed in Table~\ref{tab:table1}, and identify the reasons why only the datasets of Refs.~\cite{clark_design_2021} and~\cite{kehe_massively_2019} satisfy all requirements for applying our framework; namely, (1) enough community richness levels are sampled to clearly define at least two distinct richness tiers for comparison, and (2) enough communities are sampled within each richness tier to train and test linear regression models.

\section{Constructing the Pareto fronts -- Additional details}\label{si:pareto_details}
\subsection{Quantifying description complexity, $I(\Psi)$}\label{si:descr_compl}
The most obvious measure of model complexity is the number of model parameters. In the context of this work, this would mean setting $I(\Psi)$ (description complexity) to the number of classes (the cardinality of the partitioning). However, in the main text, we opted for a different metric, defining $I(\Psi)$ as the \textit{entropy} of the partitioning, which weighs classes differently based on the number of strains included. In this section, we comment on this choice.

\begin{figure}[h!]
    \centering
    \includegraphics[width=\linewidth]{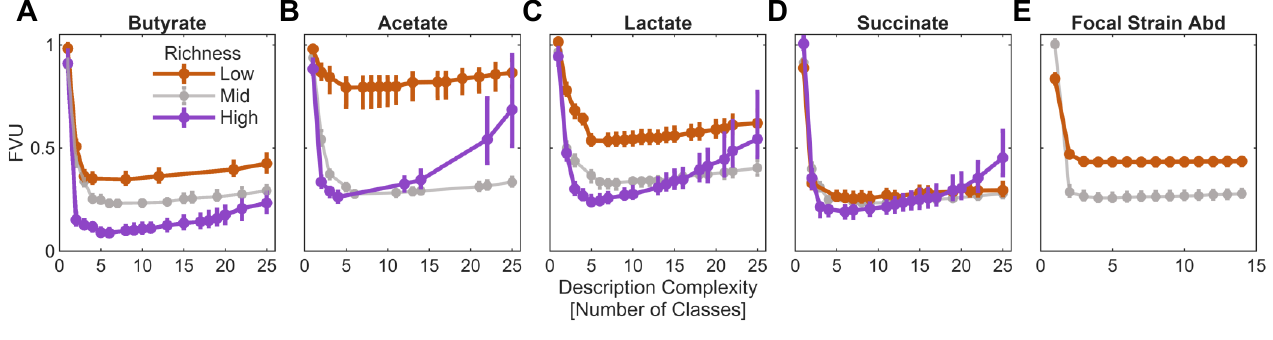}
    \caption{Redoing the analysis shown in Fig.~\ref{fig:EP} with description complexity axis $I(\Psi)$ measured in terms of number of classes of a given coarsened description $\Psi$.}
    \label{fig:num_classes}
\end{figure}

It is important to emphasize that either choice preserves the phenomenon described in this work (compare main text Fig.~\ref{fig:EP} with Fig.~\ref{fig:num_classes}; in either case, the Pareto fronts exhibit the shift indicative of emergent predictability). To explain our choice, we note that the entropy of $\Psi$ is the mutual information between the label of the coarsened class and the identity of the microscopic strain. Consider a scenario where two choices of two-group coarsenings are equally predictive, but one splits strains 50-50, while the other coarsening separates one strain from the rest of the taxa put together. If the latter is equally predictive, our definition identifies it as being simpler, and thus preferable (same predictive power, lower complexity).

This property is arguably desirable. If knowing the abundance of just one strain is sufficient to predict function, this scenario makes it easiest to gain insight into the mechanism by which community function is assured. Thus, we would prefer our framework to favor such coarsenings among others with comparable predictive power.

This definition also has some shortcomings. Most notably, any definition relying on counting strains suffers from the fact that such counting is not always straightforward, especially when some strains in the pool are close relatives. For our purposes in this work, this has not been a problem because we applied our framework to data from defined communities constructed from a well-defined set of distinct taxa. (See also the Discussion section in the main text, regarding the validity of stratifying communities by ``richness''.) In the general case, however, the hierarchical nature of relationships between strains makes the choice of ecologically-informed metrics assessing the complexity of an ecosystem description non-trivial, and a key question for future work to address.

Thus, the relative merits of our $I(\Psi)$ compared to the simple cardinality of the partitioning remain to be investigated. However, we believe this question merits further attention, and so used a non-standard $I(\Psi)$ in the main text to motivate this conversation.


\subsection{Normalization choices}\label{si:normalization}
\subsubsection{Normalizing the response variable (the community-level property $Y$)}\label{si:norm_response}
The question of quantifying the complexity of description, discussed above, is closely tied to the choice of the property of interest, and in particular, its normalization. The easiest way to see this is to observe that with our definition of $I(\Psi)$, the one-group description $\Psi_B$ (all taxa grouped together into a single category) has complexity $I(\Psi_B)=0$. Technically, this is not a zero-information description: if we think in terms of raw abundances of taxa, summing them is essentially the total community biomass, which can be informative. However, in many situations (say, if studying process rates in natural soil), properties of interest scale with biomass, and the quantity most natural to predict is the value normalized \textit{per unit biomass}. In such cases, our choice of $I(\Psi)$ is particularly appropriate. In contrast, in lab experiments where an inoculum is provided a fixed amount of resources, normalizing to biomass may not be necessary or desired, since properties of interest are meaningfully defined in the absolute sense --- in fact, the final biomass of the assembled community could itself be a property of interest.

Fortuitously, in the empirical datasets used here, the one-group coarsening (to which we colloquially refer as the ``biomass-only description'') explained almost no variance (never exceeding $15\%$ across all richness tiers and measured properties). This simplified our presentation, allowing us to declare that we were primarily focused on scenarios where the one-group coarsening explained no variance, such that we could equate $I(\Psi_B)=0$ with ``having no compositional information.'' In the interest of conciseness, in the main text it was easiest to frame this as a decision to limit consideration to only a subset of community properties -- those for which this statement is true. However, now is a good time to clarify that properties for which biomass alone explains a significant fraction of the variance are also compatible with our approach / framework. The generalization is straightforward: instead of the fraction of \textit{total} variance explained, one should consider the residual variance after regressing on the biomass. With this approach, the Pareto fronts will start at $\mathrm{FVU}=1$ \textit{by definition}, which would in fact simplify some downstream analyses. However, defining the $y$ axis of our diagrams in this way would make them appear rather less intuitive on first encounter. To summarize, limiting consideration to properties that are not explainable by biomass is a choice made for ease of presentation, not a limitation of our framework.

One minor consequence of this choice was for the computational invasion experiments (GLV \& CRM), where making this analysis comparable with our empirical curves required us to exclude the instances ($\approx 15\%$ of cases) where the combined standardized abundances could explain $>$10\% of total variance in simulated ``community invadability'' data at high richness. Had this variance been defined as the residual variance after regressing on the biomass-only prediction, applying such exclusion criteria would not have been necessary. However, the advantage offered by greater simplicity of presentation of our key concepts outweighed this consideration.

\subsubsection{Normalizing the predictor variables}
\label{si:norm_predictors}
In our analysis, the input variables supplied as ``abundance matrix'' $n$ were first standardized to zero mean and unit variance. All coarsening schemes were then applied to these standardized variables.

To understand this choice, it is important to note some limitations of the datasets. The study of Kehe \textit{et~al.}~\cite{kehe_massively_2019} did not directly measure the full composition of the final assembled communities: only the abundance of the focal strain \textit{H. Frisingense} was tracked (in arbitrary fluorescence units). The available compositional information is limited to the makeup (binary presence/absence) of the set of droplets combined by the authors' microfluidic platform to form the community initially (each droplet carrying individuals of a given strain). These binary variables are proxies for the \textit{initial} abundances of community members, and in this sense, they still satisfy the criterion of Fig.~1 of ``following a consistent assembly protocol from a known compositional state.'' However, the relationship between this binary $n$ matrix and the actual abundance values is only approximate, and includes unknown proportionality factors.

In the study of Clark \textit{et~al.}~\cite{clark_design_2021}, the composition of final assembled communities was profiled more directly, using 16S sequencing (recording relative abundances) as well as recording the $\mathrm{OD}_{600}$ for each community. However, even here, the best estimate of the absolute abundances (multiplying the relative fractions of 16S counts by the recorded community OD) is still only a proxy. The proportionality factor includes the number of 16S copies on the genome, which differs between species, and a PCR amplification bias, which is unknown and often substantial~\cite{krehenwinkel2017estimating,mclaren2019consistent}. Because of this, we adopted an approach where the input abundances were standardized to zero mean and unit variance (across samples) before training any regression models. For consistency, the same standardization procedure was also applied to Kehe \textit{et~al.} data.

\begin{figure}[h!]
    \centering    \includegraphics[width=\linewidth]{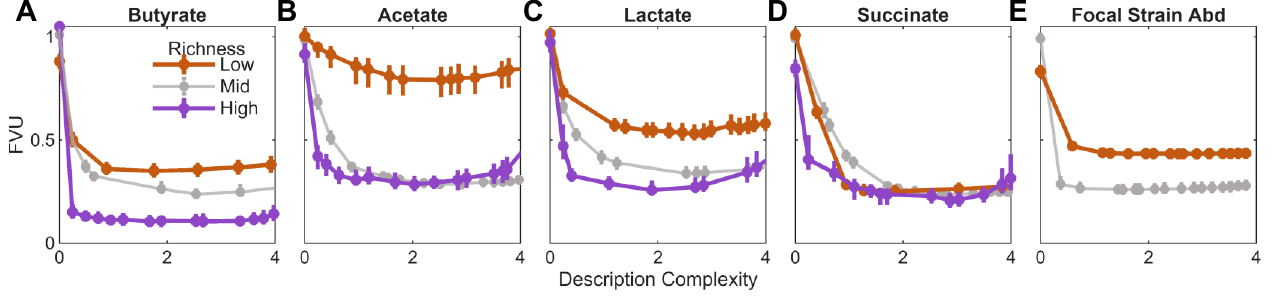}
    \caption{Reanalyzing result of Fig.~\ref{fig:EP} with raw input variables (non-standardized) used in regression models. That is, abundance data is not standardized to 0 mean and unit variance.}
    \label{fig:non_standardized}
\end{figure}

It is important to note that summing standardized abundances is not equivalent to standardizing summed abundances. However, even if absolute abundances were known, it is not clear if the biologically meaningful summation should be best implemented in units of biomass, in units of metabolic activity, or weighted in some other way. Thus, although this intuition of $\tilde n$ as the combined abundance of a putative functional class makes it easiest to explain our approach, any practical implementation on the available experimental data would implicitly include an unavoidable species-specific factor in front of every abundance value. Given this, we chose to fix these factors by simply standardizing input variables to unit variance, which is generally good practice with regression models, and came with the added benefit of improving performance. That said, we did also confirm that omitting the standardization step retains our conclusion (Fig.~\ref{fig:non_standardized}).

This standardization of abundances would certainly complicate attempts at mechanistic interpretation of the low-complexity descriptions discovered by our framework (lying along the Pareto front). In general, this strategy---discover a coarse-graining that is predictive, and then gain mechanistic insight by understanding why it works---is one of the promising applications of a framework like ours, or feature selection frameworks more generally. However, in this work, we focus on demonstrating the existence of diversity-aided emergent predictability. Interpreting the high-performance groupings is left for future work.


\subsection{Convergence of the Pareto search algorithm}
We found the convergence of our Metropolis algorithm to be most efficient if the search is performed leveraging all of the data, and evaluates a candidate grouping using its \textit{in-sample} explanatory power. Indeed, if using out-of-sample performance, one must use a very large number of cross-validation folds. Otherwise, the stochasticity of cross-validation (the variability induced by drawing different subsamplings of data) can mask true differences in performance, especially when applied to relatively small datasets like here, so that the decision whether to accept a given Metropolis move can become effectively random, impeding convergence. This large number of cross-validation folds sampled at each evaluation step would increase the runtime by orders of magnitude. Thus, in our approach, during the search phase only, performance of a candidate grouping is assessed using the in-sample FVU. After convergence, the performance of each grouping identified as belonging to the Pareto front is reassessed using out-of-sample cross-validation. This methodology was previously shown to ensure good convergence on synthetic datasets with known ground truth~\cite{zhao2024linear}, and performed well in our case as well (see section~\ref{si:robustness_test}, ``Robustness tests'').


\subsection{Training and testing of linear regression models}\label{si:cv_of_lin_reg}
As noted in the main text, we choose to focus on the simple model class of first-order linear regression for reasons stated in section~\ref{candidate_models}. This is the model class we use to apply our framework to all datasets considered in this work -- both simulated and empirical. Specifically, we write:
\[
\hat{Y}_\mu=\beta^\Psi_0+\sum_i \beta^\Psi_i \tilde n^{\Psi}_{i\mu},
\]
where $\mu$ labels communities, $\hat{Y}_\mu$ is the model's prediction of property $Y$, and $\beta_i^\Psi$ are the regression coefficients.

To assess the predictive power of a particular coarsened description $\Psi$, the data is randomly split into training and testing subsets. We note that the experimental design of both studies includes replicate communities (i.e., communities with same presence-absence combination $b_{i\mu}$ defined in section~\ref{sec:exp_setup}). We take care to keep such replicates together, such that 50\% of \textit{unique} compositions are randomly designated for model training, and the other 50\% are reserved for evaluating the out-of-sample prediction error.

Model fitting was performed using the standard ordinary least-squares method of MATLAB’s ‘fitlm’ algorithm. We then used the test set to compare model predictions to the measured values. The Fraction of Variance Unexplained (FVU) was computed in the standard way of dividing mean-squared error (MSE) of prediction by the variance of $Y$ across the test set samples. This was repeated for 100 random 50-50 splits of the data; the plots report the median value, and the error bars specify the upper and lower quartiles across these splits.

\subsection{Robustness tests}\label{si:robustness_test}
To verify that the observation of emergent predictability is robust, we perform two tests on the low-richness datasets from \cite{clark_design_2021} and \cite{kehe_massively_2019} to check that the drop in prediction error at higher richness is not due to how the data was collected or analyzed.

\begin{figure}[h!]
\centering
 \includegraphics[width=\linewidth]{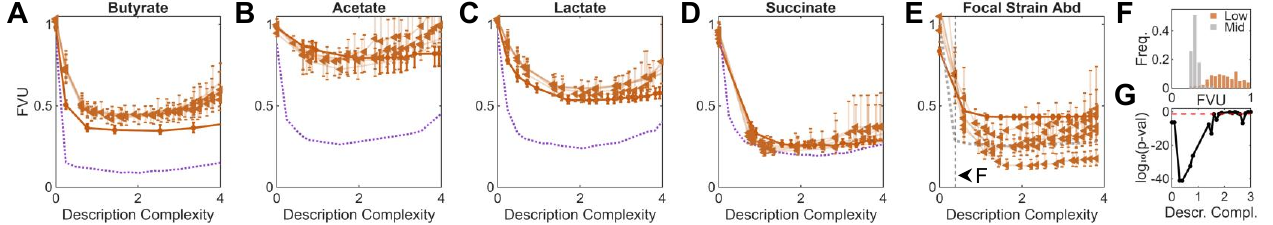}
\caption{
\textbf{Downsampling test: observed emergent predictability is robust to sample size of datasets.} To test if our observation of emergent predictability might be affected by the differnces in the number of samples per richness tier, we randomly subsample the low-richness datasets to match the total number of strain observations across the unique high-richness communities and re-obtain Pareto fronts. Plots show 5 random sub-samplings in triangles. Low-richness (orange circles) and high-richness (purple dashed line) Pareto fronts from Fig.~\ref{fig:EP} are shown for comparison. We find that for the majority of properties, the Pareto fronts obtained after downsampling do not close the gap between low richness and high richness, confirming that the shift of the Pareto front is a robust observation. The result of this test for Focal Strain Abundance (E) appears weaker than the other properties, but computing p-values from Wilcoxon rank-sum tests on the low- and mid-richness FVU distributions across description complexity (G) shows that the two Pareto fronts are statistically separable up to $\sim2$ bits. Panel F shows an example cross-section at the lowest non-zero description complexity grouping of the mid-richness Pareto, comparing the FVU distributions across cross-validation replicates that constitute the mid-richness and subsampled low-richness fronts. To perform the statistical tests, the description complexity axis was binned at a width of $0.25$ bits and FVU distributions compared within each bin. Red dashed line in G shows the $0.05$ significance threshold.
\label{fig:down_sample}}
\end{figure}

First, we perform a down-sampling test, shown in Fig.~\ref{fig:down_sample}. To test robustness to sample size, we down-sampled the low-richness communities to match the total number of strain observations across unique community combinations in the high-richness dataset. This ensures each strain is sampled for regression model fitting. We performed 5 random subsamplings of the low-richness dataset and computed the Pareto fronts for each down-sample. We then comparing these curves to the reference high-richness Pareto fronts, to check whether the subsampling might have closed the gap between them. The results are presented in Fig.~\ref{fig:down_sample}.

For all four observables of Clark \textit{et al.}, subsampling the low-richness data increases stochasticity, but does not significantly shift the Pareto front, confirming that the gap between the high- and low-richness Pareto fronts cannot be attributed to differences in dataset size.

For Kehe \textit{et al.} (Focal strain abundance), we see that the Pareto fronts computed for the downsampled data vary tremendously between subsampling replicates. To understand this, we note that the extent of subsampling is much more dramatic: the 19433 samples of the low-richness tier are downsampled to only 346 (to match the mid-richness tier); compare to 870 samples (low-richness tier) downsampled to 180 (matching the high-richness tier) for the case of Clark et al. This is compounded with the fact that for Kehe et al., the mid-richness tier is the highest available. As a result, for higher-complexity descriptions the gap either disappears, or is dwarfed by the stochasticity. We note, however, that at low-complexity end, the gap persists (see panel Fig.~\ref{fig:down_sample}F, which plots the distribution of the FVU values along the transect at 0.35 bits), confirming that emergent predictability is supported even after subsampling.

To formally confirm this, we  performed Wilcoxon rank-sum tests between the low- and mid-richness FVU distributions at a given description complexity. Panel Fig.~\ref{fig:down_sample}G plots the $p$-value of the comparison, and indicates that the high-richness Pareto front remains statistically separable from the ensemble of subsampled low-richness Pareto fronts up to $\sim2$ bits.

\begin{figure}[h!]
\centering
 \includegraphics[width=\linewidth]{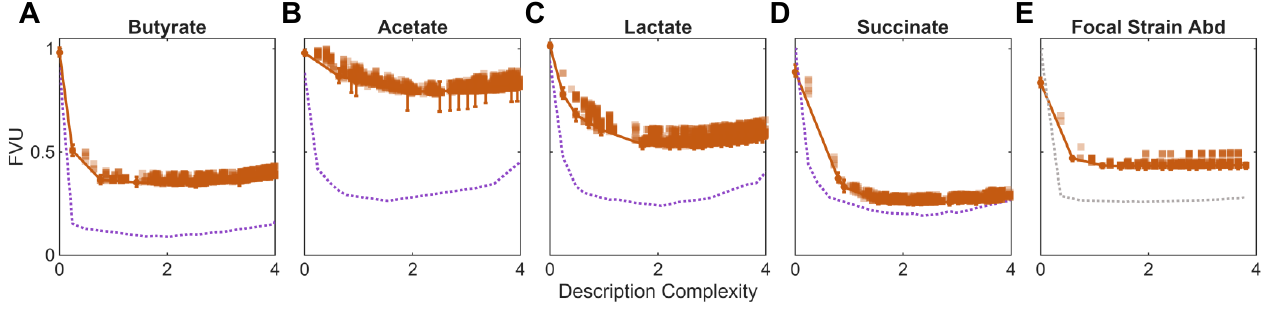}
\caption{
\textbf{Relabeling test: observed emergent predictability is robust to Pareto front detection.} To test that the emergent predictability we observe is not due to our heuristic algorithm failing to detect the Pareto front, we examine the nearest-neighbor coarsened descriptions of the low-richness Pareto front. If any neighboring coarsened descriptions provide more predictive power, possibly dropping to the low prediction error of the high-richness Pareto front, this would then indicate that emergent predictability is not robust to our analysis of the data. Scattering predictive power of perturbed coarsened descriptions making up the low-richness Pareto front, where a perturbation is the relabeling of a single strain from one group to another. Original low-richness and high-richness Pareto fronts are shown for reference. Squares are median FVU of a given relabeled coarsened description $\Psi$. Only those relabeling perturbations that fall within 20 percent of the minimum FVU are shown for visualization purposes (those with larger error are not shown). We find that for each property, no neighboring coarsened descriptions fall below the Pareto front obtained for the low-richness dataset, confirming the shift of the Pareto front is a robust observation.
\label{fig:relabel_test}}
\end{figure}

Second, we perform a relabeling test. To check how well our heuristic deduction algorithm does at inferring the Pareto front, which should be at least a local optimum, we evaluate nearby coarse-grained partitions by reassigning individual strains from their original group to all other possible groups. Fig.~\ref{fig:relabel_test} shows the best of all possible single-move relabelings relative to the low-richness and high-richness Pareto fronts presented in the main text. For all functional properties, performing such a relabeling perturbation either has no effect or decreases prediction power. This confirms that our Pareto front search algorithm does indeed converge.

\subsection{Testing for artifacts}
Checking that the emergent predictability observed in Figure~\ref{fig:EP} is not purely due to chance, we compare the result against Pareto fronts deduced after randomizing the datasets. For designing a stringent randomization test, one seeks to shuffle the data in such a way that breaks the observed effect, while preserving as much of the original structure present in the data. To apply this to the datasets from \cite{clark_design_2021} and \cite{kehe_massively_2019}, we aim to keep the abundance statistics of each species across communities fixed. For each strain in Ref.\cite{clark_design_2021}, we randomly permute its abundance in communities in which the species is initially present. Because Ref.\cite{kehe_massively_2019} only reports presence-absence of strains, we must instead shuffle by swapping the presence of each strain across a random subset of all possible community samples (i.e., not preserving presence-absence structure) while maintaining the total number of communities in which a given strain is sampled.  After doing so, we see that, as expected, the emergent predictability in high-richness communities disappears (Fig.~\ref{fig:permuteX}), as indicated by an increase in prediction error of the high-richness Pareto front obtained for the permuted abundance data. Emergent predictability is similarly broken when we instead permute the measured properties in high-richness communities, shown in Fig.~\ref{fig:permuteY}. Both tests confirm that the shift of the Pareto front with increasing community richness is not an artifact of the analysis, and that the observed emergent predictability is significant.

\begin{figure}[h!]
    \centering
    \includegraphics[width=0.5\linewidth]{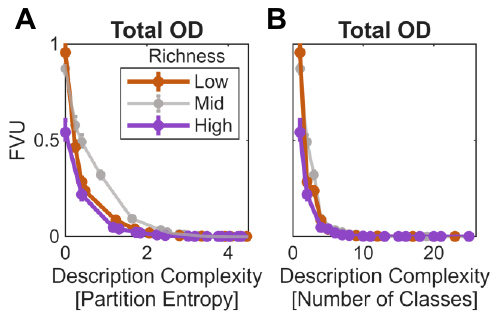}
    \caption{Using the total OD measured in the Clark \textit{et~al.} experiments as a community property for prediction, we apply our framework to produce the Pareto fronts with description complexity $I(\Psi)$ measured as partition entropy (\textbf{A}) or number of resolved classes (\textbf{B}).}
    \label{fig:odControl}
\end{figure}

\begin{figure}[h!]
\centering
 \includegraphics[width=\linewidth]{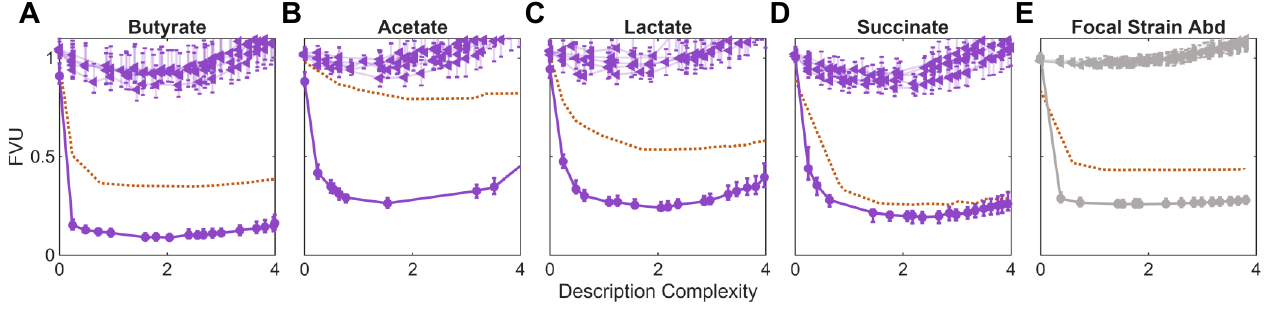}
\caption{
\textbf{Permuting abundance data breaks emergent predictability at high richness.} To test that the observation of emergent predictability is not due to an artifact in the abundance data, we perform the following permutation test to see if it breaks our observation. For Ref.~\cite{clark_design_2021} we randomize the high-richness communities by shuffling the abundances of each strain with other instances in which the strains were present (preserving the presence-absence structure of each community). For Ref.~\cite{kehe_massively_2019} we randomize the mid-richness communities by shuffling the presence of each strain while preserving the total number of samples for which a given strain is observed.  Pareto fronts are then inferred on the permuted data and plotted using triangle markers; lines with error bars show median and IQR for 5 respective random permutations. The higher-richness Pareto fronts from Fig.~\ref{fig:EP} are plotted for reference, as well as the low-richness Pareto fronts for comparison (orange dashed lines). We find that the high-richness Pareto fronts obtained from randomized abundance data do worse than the low-richness Pareto fronts for each property of interest, indicating that our permutation test breaks the observed emergent predictability and confirming the shift of the Pareto fronts is not an artifact.
\label{fig:permuteX}}
\end{figure}

\begin{figure}[h!]
\centering
 \includegraphics[width=\linewidth]{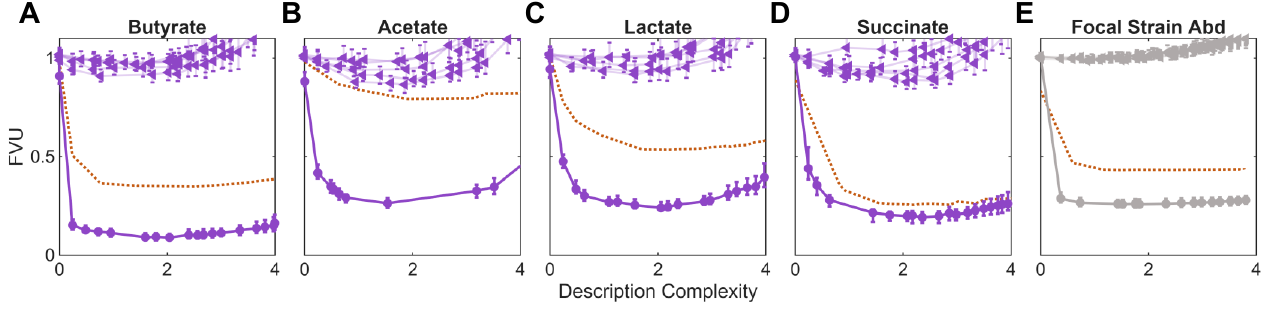}
\caption{
\textbf{Permuting property measurements breaks emergent predictability at high richness.} To test that the observation of emergent predictability is not due to an artifact in the measurements of properties of interest, we perform the following permutation test to see if it breaks our observation. Similar to Fig.~\ref{fig:permuteX}, we permuted the property data measured in high-ricnhess communities of the emprical datasets. Pareto fronts are then inferred on the permuted data and plotted using triangle markers; lines with error bars show median and IQR for 5 respective random permutations. The higher-richness Pareto fronts from Fig.~\ref{fig:EP} are plotted for reference, as well as the low-richness Pareto fronts for comparison (orange dashed lines). We find that the high-richness Pareto fronts obtained from randomized property measurements do worse than the low-richness Pareto fronts for each property of interest, indicating that our permutation test breaks the observed emergent predictability and confirming the shift of the Pareto fronts is not an artifact.
\label{fig:permuteY}}
\end{figure}

Finally, in addition to measuring fermentation end products, Clark \textit{et~al.} measured the total optical density (OD) of their communities to compute strain abundances. For completeness, Fig.~\ref{fig:odControl} presents the result of applying our framework to the OD measurements. We should emphasize, however, that community OD differs from all  other community properties used in this work, since technically, in this case the response variable and the predictor variables are \textit{not} technically independent. (Recall that for Clark et al, we use OD measurements for scaling the relative abundances profiled by 16S.) In fact, the only reason this analysis is non-trivial is because, as described above, we standardize the abundance data prior to performing the regressions (without this standardization, the total OD would trivially be a sum of input variables, and be perfectly predictable by \textit{any} coarsened representation). Thus, interpreting Fig.~\ref{fig:odControl} requires extreme caution, and we present it for completeness only. Perhaps the only possible conclusion to draw here is that ``not any set of values would be flagged by our framework as exhibiting emergent predictability,'' reinforcing the point already demonstrated by the permutation tests above.

\subsection{The raw Pareto fronts used in all comparisons}\label{si:raw_paretos}
\begin{figure}[h!]
    \centering
    \includegraphics[width=\linewidth]{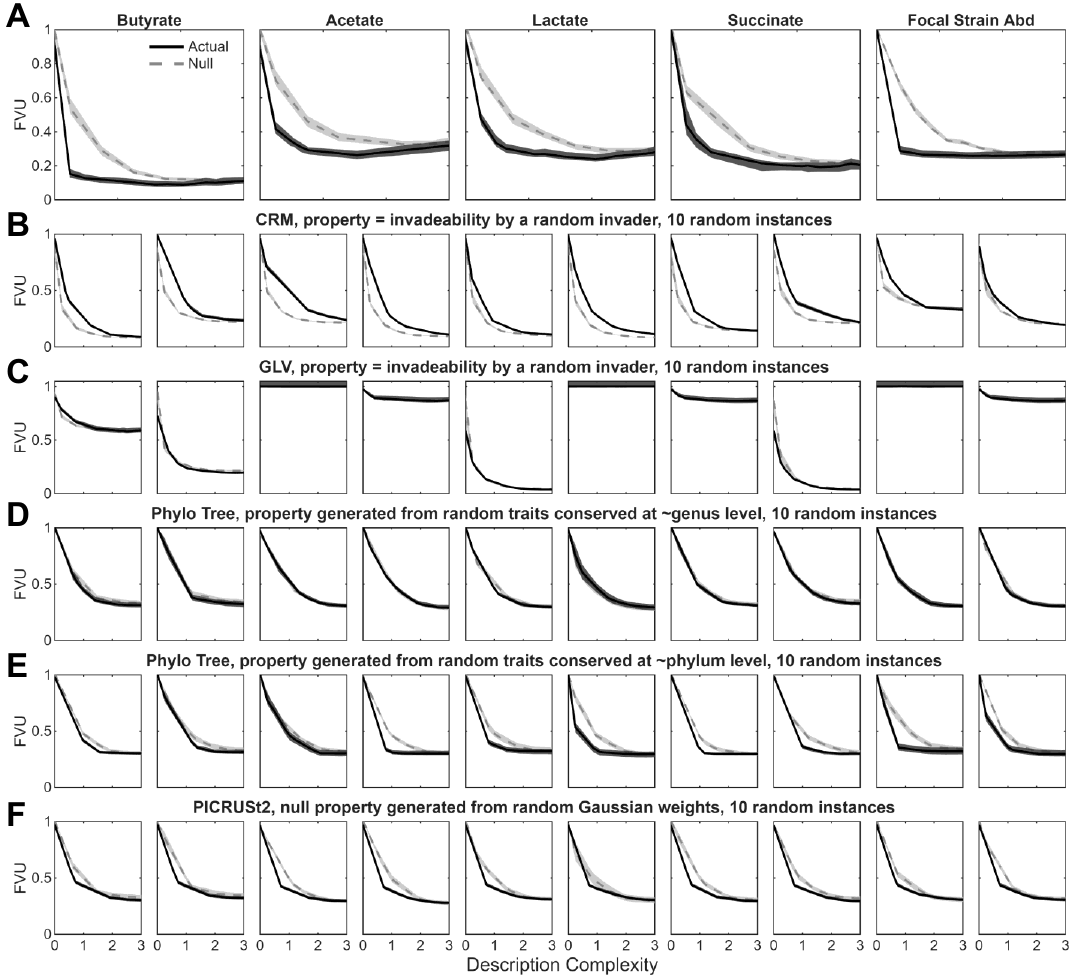}
    \caption{Plotting the raw Pareto front data for quantifying coarse-grainability in Fig.~\ref{fig:CG}. Each panel shows the median (line) and IQR (shading) across 100 cross-validation replicates for the actual Pareto versus a Gaussian null (see Methods). Each column corresponds to a different random seed for a given model instance. See text for details on simulation protocols and synthetic community property generation. Row \textbf{A}: The fermentation products measured in Ref.~\cite{clark_design_2021} and focal strain abundance measured in Ref.~\cite{kehe_massively_2019}. Rows \textbf{B,C}: Predicting post-invasion abundance of randomly drawn strains in a Consumer-Resource model (CRM) and Generalized Lotka-Volterra (GLV) model. Rows \textbf{D,E}: A synthetic community property produced from a random walk model of an evolved trait on the phylogenetic tree from Ref.~\cite{clark_design_2021}, with trait conservation depth set to the Genus- (D) or Phylum-level (E). Row \textbf{F}: A synthetic community property generated from applying PICRUSt to KEGG orthologs related to carbon metabolism in the species cultured in Ref.~\cite{clark_design_2021}. Fig.~\ref{fig:CG} is based on averaging across the 10 panels for each model as described in the Methods.
    }
    \label{fig:rawCGparetos}
\end{figure}

\begin{figure}[h!]
    \centering
    \includegraphics[width=\linewidth]{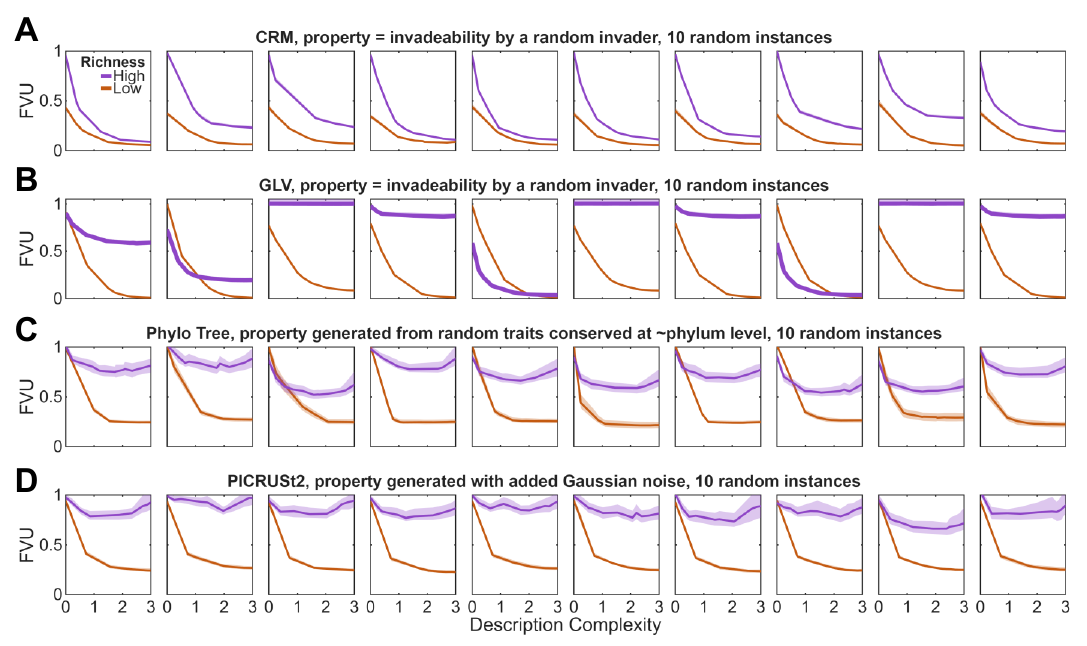}
    \caption{Plotting the raw Pareto front data for quantifying emergent predictability in Fig.~\ref{fig:EP}. Each panel shows the median (line) and IQR (shading) across 100 cross-validation replicates for the high-richness Pareto versus low-richness. Each column corresponds to a different random seed for a given model instance. See text for details on simulation protocols and synthetic community property generation. Rows \textbf{A,B}: Predicting post-invasion abundance of randomly drawn strains in a Consumer-Resource model (CRM) and Generalized Lotka-Volterra (GLV) model. Row \textbf{C}: A synthetic community property produced from a random walk model of an evolved trait on the phylogenetic tree from Ref.~\cite{clark_design_2021}, with trait conservation depth set to Phylum-level (highly structured case). Row \textbf{F}: A synthetic community property generated from applying PICRUSt to KEGG orthologs related to carbon metabolism in the species cultured in Ref.~\cite{clark_design_2021}. Fig.~\ref{fig:EP} is based on averaging across the 10 panels for each model as described in the Methods.
    }
    \label{fig:rawEPparetos}
\end{figure}

\begin{figure}[h!]
    \centering
    \includegraphics[width=\linewidth]{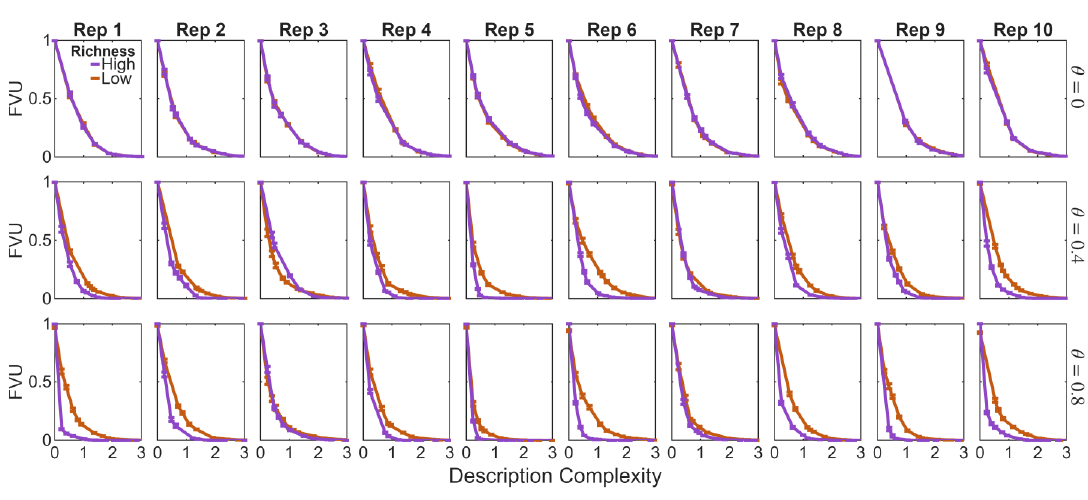}
    \caption{Plotting the raw Pareto front data for quantifying emergent predictability in the physiological feedback model presented of Fig.~\ref{fig:pH}. Each panel shows the median (line) and IQR (error bars) across 100 cross-validation replicates for the high-richness Pareto versus low-richness. Each column corresponds to a different random seed for a given model instance. See text for details on simulation protocols. Each row corresponds to a different value of feedback strength $\theta$ from the environmental variable $p$. Fig.~\ref{fig:pH} is based on averaging across the 10 panels for each $\theta$ as described in the Methods.
    }
    \label{fig:rawFeedbackParetos}
\end{figure}

Here, for more direct comparison, we show all the raw Pareto fronts found by our Metropolis search algorithm (see Methods) that were averaged for summary comparisons in the main text. Fig.~\ref{fig:rawCGparetos} shows the actual Pareto fronts versus our Gaussian null for quantifying coarse-grainability in Fig.~\ref{fig:CG}. Fig.~\ref{fig:rawEPparetos} plots the Pareto fronts for the high- and low-richness communities for models that do not exhibit emergent predictability in Fig.~\ref{fig:EP}. Fig.~\ref{fig:rawFeedbackParetos} displays the high- and low-richness Pareto fronts from the physiological feedback model, showing the 10 replicates for 3 feedback strengths that result in varying levels of emergent predictability in Fig.~\ref{fig:pH}C.


\section{Other topics}
\subsection{Structured species pool is not sufficient for emergent predictability}
\label{si:noEP_from_poolStruct}
In the main text, we explained that in any real-world setting, (at least some amount of) coarse-grainability is always expected to arise simply because species pool has structure: some species are more similar than others.

However, our proposed argument for the origin of emergent predictability may superficially appear similar (particularly when the model-derived results are rephrased in intuitive general terms in the Discussion). Consider a simple scenario with a highly structured pool of species forming two well-defined classes: A and B. Species of the same class are very similar, but A and B are distinct in important ways. Then, the following word-based argument appears plausible: ``As the number of species increases, both classes become represented by many members; thus, \textit{within} a class, the quirks of precisely which  representatives were selected will average out. In contrast, the difference \textit{between} the two classes will always persist. Thus, as richness increases, one expects the two-group description, A and B, to become increasingly predictive.'' This intuition appears to suggest that, just like coarse-grainability, emergent predictability can also arise simply as a consequence of species pool structure, contrary to our observations in the main text.

It is therefore highly instructive to convert this word-based argument into a simple quantitative model. Observing why emergent predictability is not, in fact, automatically observed in this case will not only correct a possible misconception, but will also help clarify the mechanism we propose in Fig.~\ref{fig:pH}.

As the simplest model with such a two-class species pool, we will, once again, adopt a simple ansatz where species are characterized by their contributions $\beta_i$ to community property $Y$:
\[
Y_\mu=\sum_i\beta_i n_{i\mu}\text{, where }
\beta_i = \begin{dcases}
    \beta_A +\epsilon_i, & i\in A \\
    \beta_B + \epsilon_i, & i \in B.
\end{dcases}
\]
In other words, the contribution is defined primarily by the class identity ($\beta_A$ vs. $\beta_B$), while $\epsilon_i$ (the ``quirks'') represent the small deviations of individual strains, and are assumed small. We will take strain $n_{i\mu}$ and $\epsilon_i$ to be independent and Gaussian distributed: $n_{i\mu}\sim\mathcal{N}(m_n,\sigma_n^2)$ and $\epsilon_i\sim\mathcal{N}(0,\sigma_\epsilon^2)$, with $\sigma_\epsilon^2\ll(\beta_A-\beta_B)^2$.

The bimodal distribution of coefficients $\beta_i$ means that $Y$ will exhibit strong coarse-grainability: the natural two-group description: $n_A=\sum_{i\in A} n_i$ and $n_B=\sum_{i\in B} n_i$ has a much lower error than could be expected by chance. We will now consider whether $Y$ also exhibits emergent predictability. For this, we need to compute the fraction of variance of $Y$ explained by the two-group description, and evaluate its dependence on community richness $S=S_A+S_B$. Here $S_A$ and $S_B$ denote the the number of strains in the two groups. For simplicity, we will assume that $S_A=S_B=S/2$, though this can of course be generalized.

Note that the parameter $m_n$ can itself depend on $S$ -- for example, if community is assembled under a common resource limitation, one may expect $m_n\sim1/S$; similarly, $\sigma_n$ may or may not scale with $S$ as well. For our argument below, we will only need to consider their \textit{relative} scaling. The ratio $\sigma_n/m_n$ represents the ``noisiness'' of each individual species abundance. It would be extremely unusual to see this ratio decrease with $S$. Indeed, if some mechanism ensured that the abundance of each individual species becomes increasingly predictable with diversity, this would directly lead to an extremely strong form of emergent predictability, whereby not only the macroscopic community-level properties, but even the microscopic-level description is increasingly predictable. No empirical indications support such a strong form. Thus, in general we expect $\sigma_n/m_n\sim S^\alpha$ with $\alpha\ge0$. Our argument below will not require any stronger assumptions; this general statement will be sufficient.

Computing the variance of the community property $Y$, we can separate the total variance into two contributions:
\[
\mathrm{var}_\mu[Y_\mu] = \underbrace{\mathrm{var}\left[ \sum_{i\in A} n_{i\mu} \beta_{A} + \sum_{i\in B} n_{i\mu} \beta_{B}\right]}_{\mathrm{var}_\mathrm{coarse}} + \underbrace{\mathrm{var}\left[ \sum_i n_{i\mu}\epsilon_i\right]}_{\mathrm{var}_{\mathrm{quirks}}},
\]
where the first term is precisely the variance explainable by the coarse-grained description, and the second term is the residual contribution from the individual ``quirks.'' To test for emergent predictability in $Y$, we need to compute the fraction of the explainable variance relative to the total. Simplifying each term we have
\[
\begin{aligned}
    \mathrm{var}_{\mathrm{coarse}} &= (S_A \beta_A^2  + S_B \beta_B^2) \sigma_n^2 = \frac 12 S\sigma_n^2\left(\beta_A^2+\beta_B^2\right)\\
    \mathrm{var}_{\mathrm{quirks}} &= (S_A + S_B)\sigma_\epsilon^2(\sigma_n^2+m_n^2) = S\sigma_n^2\sigma_\epsilon^2(1+(\dots)S^{-2\alpha})\simeq S\sigma_n^2 C \sigma_\epsilon^2
\end{aligned}
\]
In the last line, we used the fact that $\alpha\ge0$, and therefore the $S^{-2\alpha}$ term in the brackets is either negligible, or at most provides a constant correction of order 1. We conclude that in this example, both contributions to the variance scale in the exact same way with $S$, and the fraction of variance left unexplained by the two-group model remains constant:
\[
 \frac{\mathrm{var}_{\mathrm{quirks}}}{\mathrm{var}_{\mathrm{coarse}}+\mathrm{var}_{\mathrm{quirks}}} = \frac{2C\sigma_\epsilon^2}{ \beta_A^2+\beta_B^2+2C\sigma_\epsilon^2}.
\]
Therefore, the observable $Y$ exhibits no emergent predictability, despite being strongly coarse-grainable due to species pool structure. Intuitively, the individual strain ``quirks'' self-average with richness -- but so does the variance explained by the coarse model, scaling in the exact same way. Thus, in the \emph{absolute} sense, the two-class model is an increasingly good one, and this is the phenomenon of self-averaging (unexplained variance decreases). However, the unexplained \emph{fraction} of variance remains constant (no emergent predictability).
	
Now, imagine if some mechanism were to maintain a finite range of variability along the A--B axis, such that some communities assembled into mostly A-type strains, while others assembled into mostly B-type strains. Then, as the $\epsilon_i$ average out, the variance explained by the class representation would indeed become an increasingly dominant fraction. This is what our Fig.~5 model achieves. In a pH-based intuition: the existence of two classes, acid lovers and acid haters, is not enough. But if there is a feedback that drives community pH in either direction based on initial composition, such that the community alternately assembles into a low-pH state dominated by acid-lovers, or a high-pH state dominated by acid haters -- this community will display emergent predictability.

In the actual model we use in Fig.~5, note that turning the $p$-dependent feedback off makes the metabolic activity $a_i$ identical for all species at all times; all species ``grow as one'' until the resource is depleted. Then, it is trivial to see that the contributions to both $p$ and $Y$ experience pure self-averaging, which affects both quantities in the exact same way. In this limit, the failure to observe emergent predictability has the same origin as in the model just described.

\subsection{PCA on Clark~\textit{et~al.} functional space}
\begin{figure}[h!]
    \centering
    \includegraphics[width=0.4\linewidth]{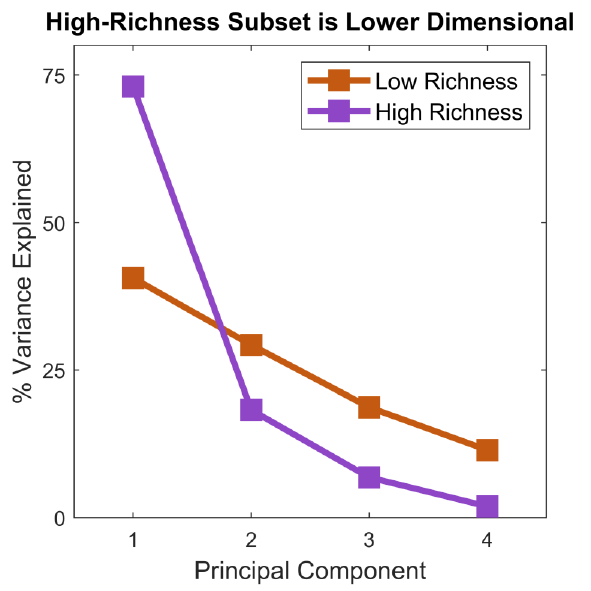}
    \caption{PCA spectra of the four fermentation byproduct concentrations measured in Ref.~\cite{clark_design_2021}, computed separately in the two tiers of community richness. For high-richness, the dominant component explains $73\%$ of variance, contrasted with $41\%$ for low-richness.}
    \label{fig:si_fig:pca}
\end{figure}

The experiments of Ref.~\cite{clark_design_2021} measured the final concentration of four fermentation byproducts in each sample community. We can view these measurements as points in a four-dimensional ``functional space.'' In Fig.~\ref{fig:pH}A, we perform principal component analysis on this functional property data and plot the projections along the first two principal axes for the measurements in low- versus high-richness communities, showing that at higher richness community function is spanned by a lower-dimensional subspace. Fig.~\ref{fig:si_fig:pca} quantifies this by plotting the variance explained by each principal component. The plot demonstrates that as richness increases, the first principal axis explains an increasing fraction of total variance, and confirms the numbers quoted in the main text.

\section*{Supplementary figures}

\clearpage
\section*{Supplementary tables}
\begin{table}[h!]
    \centering
    \begin{tabular}{||p{1.5cm}|p{2.1cm}|p{3cm}|p{4cm}|p{3cm}||}
        \hline
        Ref. & Context & $\geq$2 richness tiers? & \#(samples) / \#(strains) & Satisfies criteria? \\ [0.5ex]
        \hline\hline
        \cite{clark_design_2021} & microbial & yes & 870/25 (low)\par 477/25 (mid)\par 180/25 (high) & yes \\
        \hline
        \cite{kehe_massively_2019} & microbial & yes & 19433/14 (low)\par 346/14 (mid) & yes\\
        \hline
        \cite{faith2014identifying} & microbial & yes & 21/17 (low)\par 30/17 (mid) & not enough data\\
        \hline
        \cite{jungers2021diversifying} & grasslands & yes & 350/31 (low)\par 200/31 (mid)\par 50/31 (high) & not enough data; not all strains sampled at low and mid richness\\
        \hline
        \cite{hector1999plant,spehn2005ecosystem} & grasslands & yes & 345/32 (low)\par 22/32 (mid) & not enough data at mid richness\\
        \hline
        \cite{weigelt2010jena,roscher2005overyielding} & grasslands & yes & 200/60 (low)\par 40/60 (mid)\par 10/60 (high) & not enough data\\
        \hline
        \cite{reich2001plant,reich2004species,reich2009elevated} & grasslands & no & Not evaluated & not enough data; no richness tiers\\
        \hline
        \cite{van2003positive} & grasslands & yes & Not evaluated & plots either had all or one species\\
        \hline
        \cite{isbell2011increasing} & grasslands & no & Not evaluated & low richness only\\ [1ex]
        \hline
    \end{tabular}
    \caption{Catalog of published datasets that assemble communities from a fixed library of species/strains.}
    \label{tab:table1}
\end{table}

\end{document}